\documentclass{JHEP3} 

\usepackage{amsfonts}
\usepackage{amsmath}
\usepackage{amssymb}
\usepackage{comment}
\usepackage{epsfig,multicol}
\usepackage{fancyhdr}
\usepackage{float}
\usepackage{graphicx}
\usepackage{mathrsfs}
\usepackage{subfigure}
\usepackage[small,bf]{caption}
\usepackage{caption} 
\newcommand\fverb{\setbox\fverbbox=\hbox\bgroup\verb}
\newcommand\fverbdo{\egroup\medskip\noindent%
			\fbox{\unhbox\fverbbox}\ }
\newcommand\fverbit{\egroup\item[\fbox{\unhbox\fverbbox}]}
\newbox\fverbbox

\newcommand{\beqa}{\begin{eqnarray}}
\newcommand{\eeqa}{\end{eqnarray}}
\newcommand{\CR}{\nonumber \\}

\newcommand{\beq}{\begin{equation}}
\newcommand{\eeq}{\end{equation}}
\newcommand{\bea}{\begin{eqnarray}}
\newcommand{\eea}{\end{eqnarray}}

\newcommand{\CL}{{\mathcal L}}

\newcommand{\CN}{{\mathcal N}}
\newcommand{\CO}{{\mathcal O}}

\newcommand{\CS}{{\mathcal S}}

\newcommand{\CW}{{\mathcal W}}

\newcommand\SO{{\rm SO}}

\newcommand{\nt}{\nonumber\\}
\newcommand{\nn}{\nonumber}

\newcommand{\cW}{{\cal W}}

\newcommand{\hc}{{\hat c}}

\newcommand{\bbC}{{\mathbb C}}

\newcommand{\ba}{\begin{eqnarray}}
\newcommand{\ea}{\end{eqnarray}}

\newcommand{\back}{\!\!\!\!\!\!}

\newcommand{\parfrac}[2]{\frac{\partial #1}{\partial #2}}

\newcommand{\alp}{\alpha}

\newcommand{\lam}{\lambda}
\newcommand{\eps}{\epsilon}

\newcommand{\cdotsb}{\cdots{\:\!\!}}

\newcommand{\Wi}{W^{(i)}}
\newcommand{\Wj}{W^{(j)}}
\newcommand{\Li}{L^{(i)}}

\newcommand{\wtRn}{\widetilde{R_n}}

\newcommand{\valp}{{\vec\alpha}}
\newcommand{\vrho}{{\vec\rho}}
\newcommand{\vlam}{{\vec\lam}}
\newcommand{\vvphi}{{\vec\varphi}}
\newcommand{\ve}{{\vec e}}
\newcommand{\vu}{{\vec u}}
\newcommand{\vtheta}{{\vec\theta}}
\newcommand{\vw}{{\vec w}}

\newcommand{\tm}{{\tilde m}}

\newcommand{\mh}{{\hat m}}
\newcommand{\Mh}{{\hat M}}

\input{colordvi.tex}

\title{$\CW_3$ irregular states and isolated $\mathcal{N}=2$  superconformal field theories}

\preprint{KEK-TH-1595, RIKEN-MP-66}

\author{
Hiroaki Kanno$^{\spadesuit}$\footnote{kanno@math.nagoya-u.ac.jp}\,\,, 
Kazunobu Maruyoshi$^{\heartsuit}$\footnote{maruyosh@caltech.edu}\,\,, 
Shotaro Shiba$^{\diamondsuit}$\footnote{sshiba@post.kek.jp}\,\,,
Masato Taki$^{\clubsuit}$\footnote{taki@riken.jp}
\vspace*{0.5cm}\\
\llap{${}^{\spadesuit}$}
Graduate School of Mathematics and KMI, Nagoya University,\\{\:\!}
Nagoya, 464-8602, Japan
\\
\llap{${}^{\heartsuit}$}
California Institute of Technology,\\{\:\!}
452-48, Pasadena, California 91125, USA
\\
\llap{$^{\diamondsuit}$}
High Energy Accelerator Research Organization (KEK),\\{\:\!}
1-1 Oho, Tsukuba, Ibaraki 305-0801, Japan
\\
\llap{$^{\clubsuit}$}
Mathematical Physics Lab., RIKEN Nishina Center, Saitama 351-0198, Japan
}

\abstract{
We explore the proposal that the six-dimensional $(2,0)$ theory on the Riemann surface 
with irregular punctures leads to a four-dimensional gauge theory coupled to
the isolated $\mathcal{N}=2$ superconformal theories of Argyres-Douglas type, 
and to two-dimensional conformal field theory with irregular states. 
Following the approach of Gaiotto-Teschner for the Virasoro case,
we construct $\CW_3$ irregular states by colliding a single $SU(3)$
puncture with several regular punctures of simple type.
If $n$ simple punctures are colliding with the $SU(3)$ puncture,
the resulting irregular state is a simultaneous eigenvector
of the positive modes $L_n, \ldots, L_{2n}$ and $W_{2n}, \ldots, W_{3n}$
of the $\CW_3$ algebra. 
We find the corresponding isolated SCFT with an
$SU(3)$ flavor symmetry as a nontrivial IR fixed point on the Coulomb branch of the 
$SU(3)$ linear quiver gauge theories, 
by confirming that its Seiberg-Witten curve 
correctly predicts the conditions for the $\CW_3$ irregular states.
{We also compare these SCFT's with} the ones obtained from the BPS quiver method.
}

\begin{document}

\section{Introduction}
\label{sec:intro}

The twisted compactification of the six-dimensional $\mathcal{N} =(2,0)$ theory
on a punctured Riemann surface $C_{g,n}$ gives rise to a large class of $\mathcal{N} =2$ 
superconformal field theories (SCFT) in four dimensions~\cite{Gaiotto:2009we, Gaiotto:2009hg},
{called class $\mathcal{S}$}. 
The gauge couplings of these theories are exactly marginal and the space of the gauge coupling 
is identified with the moduli space $\mathcal{M}_{g,n}$ of the complex structure of $C_{g,n}$,
on which the $S$-duality group acts. 
There is another class of SCFT's which is 
isolated in the sense that they do not allow marginal deformations. This class of
theories was originally found as a nontrivial IR fixed point on the Coulomb branch of 
asymptotically free gauge theories and called Argyres-Douglas type 
\cite{Argyres:1995jj, Argyres:1995xn, Eguchi:1996vu}. 
The characteristic feature of these theories is that mutually non-local BPS particles 
get massless at the superconformal point. 
Recently, it was shown that this class of SCFT's can also be constructed by the compactification of 
the six-dimensional $\CN=(2,0)$ theory on a sphere with an irregular puncture \cite{Cecotti:2010fi, Cecotti:2011rv}.

From the six-dimensional viewpoint a remarkable correspondence has been uncovered
\cite{Alday:2009aq, Wyllard:2009hg}:
the instanton partition function \cite{Nekrasov:2002qd} of the four-dimensional $\mathcal{N} =2$
gauge theory of class $\mathcal{S}$ 
is exactly equal to the conformal block on $C_{g,n}$ of $\CW$ algebra in two dimensions, 
with a suitable identification of the parameters.
Then, an extension of the correspondence to isolated SCFT's has been proposed 
in \cite{Bonelli:2011aa, Gaiotto:2012sf}
by finding that the two-dimensional CFT counterpart of the irregular puncture
is an irregular state which is a simultaneous eigenstate of the higher Virasoro generators.
In \cite{Gaiotto:2012sf} the irregular state has been constructed 
by a collision (or confluence) of several Virasoro vertex operators corresponding to the regular punctures.
Similar construction was given by using the matrix model in \cite{Nishinaka:2012kn, Rim:2012tf}.
In this article we explore this proposal for the irregular states of $\CW_{3}$ algebra
and isolated SCFT's with an $SU(3)$ flavor symmetry.

To the compactification on $C_{g,n}$ of $\mathcal{N} = (2,0)$ theory of $A_N$ type, 
one can associate the Hitchin system on $C_{g,n}$ with gauge group 
$SU(N+1)$~\cite{Gaiotto:2009hg, Bonelli:2009zp, Nanopoulos:2009uw}.
The Seiberg-Witten curve of the four-dimensional theory is identified with 
the spectral curve of the Hitchin system. At a regular puncture the $\mathfrak{sl}(N+1)$ 
valued holomorphic one-form\footnote{Due to the twisted compactification
the field $\varphi(z)$ becomes a one-form on $C_{g,n}$.} 
$\varphi(z)$ of the Hitchin system has a simple pole and the residue is associated with mass parameters. 
If $\varphi(z)$ has a pole of higher order, the puncture is called irregular. 
The coefficients of the spectral curve: $\mathrm{det}\,(x - \varphi(z)) = x^{N+1} + \phi_2(z) x^{N-1} +
\cdots + \phi_{N} (z) x + \phi_{N+1} (z) =0$ gives a $j$-th differential  $\phi_j(z)$.
The parameter $x$ in the spectral curve is a fiber coordinate of the cotangent bundle 
$T^{*} C_{g,n}$  and the Seiberg-Witten differential is the pull-back of the canonical one-form
$\lambda = x dz$ on $T^{*} C_{g,n}$ to the spectral curve.

The AGT correspondence \cite{Alday:2009aq, Wyllard:2009hg} tells us 
that the \lq\lq expectation value\rq\rq\ $\langle W^{(j)}(z) \rangle$ of the spin $j$ current 
in the $\mathcal{W}_{N+1}$ algebra gives the $j$-th differential $\phi_j(z)$.
Now at an irregular puncture the $j$-th differential $\phi_j(z)$
has a pole of the order higher than $j$. 
Since the spin $j$ current is expanded as $W^{(j)}(z) = 
\displaystyle{\sum_{n \in \mathbb{Z}}}~W^{(j)}_n (z - a)^{-j-n}$ around a puncture $z=a$,
this implies that some of the positive modes $W^{(j)}_n$ do not annihilate 
the state associated with the irregular puncture. 
Namely it is not a primary state any more.
By using such irregular states, to any Riemann surface with irregular punctures 
we can construct the irregular conformal block, as is the case with regular punctures. 
We note that the irregular conformal block also appears in connection with
the so-called confluent KZ equations
~\cite{MR2451670, MR2812339, MR2735220}.

The isolated SCFT in four dimensions has several (off-critical) deformation parameters 
from the superconformal fixed point on the Coulomb branch: the VEV's of relevant deformation operators $v_{i}$
paired with the corresponding couplings $c_i$, and mass parameters.
The parameters $v_{i}$ can be considered as the Coulomb moduli of the isolated SCFT.
We can incorporate the relevant parameters in the Seiberg-Witten
curve as the coefficients of the Laurent expansion of the $j$-th differential $\phi_j(z)$
around the pole of higher degree. This is the reason why we need
irregular singularities for the Seiberg-Witten geometry of the isolated SCFT.

This, however, indicates also that irregular singularities do not necessarily lead to the isolated SCFT.
Namely, it is easy to see that when the singularity of the differential is too mild,  
there is no room to include the above-mentioned deformation parameters.
This case simply corresponds to an asymptotically-free gauge theory with a Lagrangian description, {\it e.g.,}
$SU(N+1)$ pure super Yang-Mills (SYM) theory.
In the context of the AGT correspondence, these {\it milder} irregular states were defined as 
a coherent (Whittaker) state in the Verma module~\cite{Gaiotto:2009ma, Marshakov:2009gn}.
This has been generalized to several cases: we can find the defining conditions for 
such states in the Verma module of the chiral algebra of 
the corresponding CFT~\cite{Taki:2009zd, Kozcaz:2010yp, Wyllard:2010rp, Wyllard:2010vi, Kanno:2011fw, 
Belavin:2011pp, Bonelli:2011jx, Ito:2011mw, Wyllard:2011mn, Keller:2011ek, Kanno:2012xt}. 

In this paper, we consider the {\it wilder} irregular states of $\CW_{3}$ algebra 
which correspond to the isolated SCFT's with an $SU(3)$ flavor
symmetry, extending the $SU(2)$ case discussed in previous literatures.
After reviewing the $SU(2)$ case in section \ref{sec:Virasoro},
we introduce in section \ref{sec:W3} 
an irregular state $\vert I_n \rangle$ of $\CW_3$ algebra by taking an appropriate limit 
of colliding $(n+1)$ punctures. 
For $SU(3)$ we have two types of regular punctures; puncture of simple type and of full type. 
In this paper we only consider the case where $n$ simple punctures are colliding with
a single puncture of full type, leaving other possibilities for
future investigation. The $\CW_3$ algebra consists of the 
energy momentum tensor $T(z) = \displaystyle{\sum_{n \in \mathbb{Z}}}~L_n z^{-2-n}$
and the spin-3 current  $W(z) = \displaystyle{\sum_{n \in \mathbb{Z}}}~W_n z^{-3-n}$.
Using $\CW_3$ Ward identities for the primary
states associated with the regular punctures, we derive the characterizing
conditions for the irregular state $\vert I_n \rangle$. It turns out that $\vert I_n \rangle$
is a simultaneous eigenstate of $L_{n}, \ldots, L_{2n}$ and $W_{2n}, \ldots, W_{3n}$
and annihilated by higher modes $L_{k > 2n}$ and $W_{\ell > 3n}$.

The gauge theory counterpart will be analyzed in section \ref{sec:SU(3)}, 
after considering the simpler $SU(2)$ case in section \ref{sec:SU(2)}.
The two-dimensional CFT analysis implies that if we put the irregular state $\vert I_n \rangle$ at $z=0$,
the corresponding Seiberg-Witten curve of $SU(3)$ gauge theory is $x^3 + \phi_2(z) x + \phi_3(z) =0$
where the quadratic differential $\phi_2(z)$ and the cubic differential $\phi_3(z)$ have
a pole of order $2n+2$ and $3n+3$, respectively. 
We show that an isolated SCFT with
such singularity arises in a scaling limit of $SU(3)$ linear quiver gauge 
theory which is obtained by the compactification of $\CN =(2,0)$ theory
on the Riemann sphere with regular punctures. 
By the scaling limit we make the punctures other than at infinity colliding at the origin. 
We conclude with several discussions in section \ref{sec:conclusions}.

The conventions of $\CW_{3}$ algebra and the $A_{2}$ Toda theory are fixed in Appendix \ref{sec:Toda}.
In Appendix \ref{sec:BMTGT}, we will see that depending on the convention of the basis of the Verma module,
it is possible to derive two different conditions for the irregular state.
In Appendix \ref{sec:U(1)}, we summarize the fact that the irregular state for the $U(1)$ current algebra
is a familiar coherent state in the Fock space of free boson.


\section{Irregular states of Virasoro algebra}
\label{sec:Virasoro}
  The six-dimensional $\CN=(2,0)$ theory of type $A_{N}$ on a Riemann surface $C_{g,n}$,
  allowing only regular punctures, with a suitable twist 
  gives a class of $\CN=2$ superconformal field theories in four dimensions.
  Let us denote this $\CN=2$ theory by $\CS(A_{N}, C_{g,n})$.
  The regular puncture comes from the codimension-two defect of the six-dimensional theory 
  and is classified by a Young diagram with $N+1$ boxes \cite{Gaiotto:2009we} 
  including the information of a flavor symmetry.
   In this paper we only consider $N=1$ and $2$ cases.

  The AGT correspondence \cite{Alday:2009aq}
  (and generalization to higher rank case \cite{Wyllard:2009hg, Mironov:2009by}) 
  relates the Nekrasov instanton partition function
  of $\CS[A_{N}, C_{g,n}]$ on the Omega background $(\epsilon_{1}, \epsilon_{2})$ 
  with the conformal block of $\CW_{N}$ algebra on $C_{g,n}$.
  We should note that in this correspondence we need to specify a marking of $C_{g,n}$.
  On the gauge theory side this leads to a particular weak coupling description, while
  on the CFT side this is necessary to compute the conformal block.
  A simple example of the correspondence is the case with $N=1$ and $C_{0,4}$.
  In this case $\CS[A_{1}, C_{0,4}]$ is an $SU(2)$ gauge theory with four fundamental hypermultiplets
  and the AGT correspondence relates the Nekrasov partition function $Z_{\CS[A_{1}, C_{0,4}]}$ with
  the conformal block $\mathcal{B}[C_{0,4}]$ of the Virasoro algebra on four-punctured sphere;
    \begin{align}
    Z_{\CS[A_{1}, C_{0,4}]}
     =     \mathcal{B}[C_{0,4}].
           \label{originalAGT}
    \end{align}
  The $\CN=2$ theory $\CS[A_{1}, C_{0,4}]$ has vanishing beta function
  leading to the superconformal invariance at the origin of the Coulomb moduli space 
  and vanishing hypermultiplet mass parameters.
  This is simply due to the fact that there only appear the regular punctures.
  
  In this paper we will investigate the AGT correspondence for asymptotically free gauge theories, 
  especially for the isolated SCFT's appearing as a nontrivial IR fixed point on the Coulomb moduli space 
  of $\CN=2$ quiver gauge theory.
  This extension requires an insertion of {\it irregular} punctures on the Riemann surface 
  so that the Seiberg-Witten differential has higher singularities.
  We explain how these irregular punctures appear on the both sides of the correspondence.
  In this section, we provide a brief introduction to such extensions in the $N=1$ case, 
  namely Virasoro algebra and $SU(2)$ gauge theory.


\subsection{Irregular states and asymptotically free gauge theories}
  Let us denote the Riemann surface of genus $g$ with $n$ regular punctures 
  and $\ell$ irregular punctures by $C_{g,n,\{d_{i}\}}$ where $i=1,\ldots,\ell$ and $d_{\ell}$ are degrees of 
  irregular punctures.
  There is only one type of regular punctures in the $A_{1}$ case and
  each puncture is associated with an $SU(2)$ flavor symmetry.
  To this Riemann surface we have a four-dimensional gauge theory $\CW(A_{1}, C_{g,n,\{d_{i}\}})$.
  When $\ell=0$ we denote the Riemann surface as $C_{g,n, \{ d_{i} \} =\emptyset} \equiv C_{g,n}$, 
  which reduces the theory to the class $\CS$ of SCFT's.
  The presence of irregular punctures changes the theory into asymptotically free.
  The matter content of the theory is determined by the degree of the irregular punctures.
  
  The simplest example with irregular singularities is the pure $SU(2)$ super Yang-Mills (SYM) theory which is associated 
  with $C_{0,0,\{ \frac{3}{2}, \frac{3}{2} \}}$, namely a sphere with two irregular punctures of degree $3/2$.
  Indeed, the Seiberg-Witten curve of $SU(2)$ SYM theory with the Coulomb moduli $u$ and the 
  dynamical scale $\Lambda$ is written as
    \bea
   x^{2} = \phi_{2}(t)
     =     \frac{\Lambda^{2}}{t^{3}} + \frac{u}{t^{2}} + \frac{\Lambda^{2}}{t},
    \eea
  where the Seiberg-Witten differential is $\lambda_{{\rm SW}} = x dt$  \cite{Gaiotto:2009hg, Gaiotto:2009ma}.
   Counting the degrees of punctures with respect to the differential $\lambda_{{\rm SW}}$, we see 
  that the theory is associated with $C_{0,0,\{ \frac{3}{2}, \frac{3}{2} \}}$ 
  whose punctures are at $t=0, \infty$.
  To add one hypermultiplet changes the degree of one of the irregular punctures as follows:
    \bea
    \phi_{2}
     =     \frac{\Lambda^{2}}{4t^{4}} + \frac{\Lambda m}{t^{3}} + \frac{u}{t^{2}} + \frac{\Lambda^{2}}{t},
           \label{Nf=1}
    \eea
  where $m$ is the mass parameter of the hypermultiplet.
  The irregular puncture at $t=0$ now has degree $2$ and
  the theory is associated with $C_{0,0,\{ 2, \frac{3}{2} \}}$.
  
  The decoupling of the $SU(2)$ gauge group in the above examples leads to a sphere 
  with one irregular and one regular puncture; $C_{0,1,\{ \frac{3}{2} \}}$ or $C_{0,1,\{ 2 \}}$.
  In other words, $\CW(A_{1}, C_{0,1,\{ \frac{3}{2} \}})$ and $\CW(A_{1}, C_{0,1,\{ 2 \}})$
  are the theory of ``no hypermultiplet'' and of two free hypermultiplets respectively.
  These two types of two-punctured sphere indeed exhaust possible choices 
  to have an asymptotically free $SU(2)$ gauge theory with a Lagrangian description. 
  The other types of two-punctured sphere where the degree of the irregular puncture is higher than 2 
  lead to isolated SCFT's which do not have Lagrangians.
  
  For the above two cases which allow a Lagrangian description, 
  the corresponding states on the two-dimensional CFT side were found in \cite{Gaiotto:2009ma}.
  We demonstrate the idea by reducing the number of flavors 
  by one out of the original AGT correspondence \eqref{originalAGT}.
  The starting point on the CFT side is the conformal block  $\mathcal{B}[C_{0,4}]$.
  Let us introduce the following state made from two primaries:
\begin{align}
V_{\Delta_2}(z)
\vert \Delta_1 \rangle\vert_{\Delta}
\propto
\sum_{Y,Y'}
\frac{\vert \Delta,Y \rangle\,
Q
\left(\Delta\right)
^{-1}_{Y,Y'}
\langle \Delta,Y'\vert V_{\Delta_1}(z)\vert \Delta_1\rangle}
{\langle \Delta\vert V_{\Delta_1}(z)\vert \Delta_1\rangle}
=:\vert  \widetilde{R} (\Delta_2, \Delta_1 ; z)\rangle,
\end{align}
where $\vert_\Delta$ is the projection onto the Verma module $\mathcal{V}_\Delta$.
The corresponding projector is $1_\Delta=\sum_{Y,Y'}\vert \Delta,Y \rangle\,
Q
\left(\Delta\right)
^{-1}_{Y,Y'}
\langle \Delta,Y'\vert $,
where the summation is over two Young diagrams (partitions) $Y$ and $Y'$. 
The descendants $\vert \Delta,Y \rangle=L_{-Y}\vert \Delta\rangle$ span the Verma module
and $Q_{Y,Y'}=\langle\Delta, Y\vert \Delta,Y' \rangle$ is the Kac-Shapovalov matrix 
which is assumed to be non-degenerate.
The state on the right hand side is of course a regular vector $\vert \widetilde{R}\rangle\in\mathcal{V}_\Delta$ in the module
and the leading term of the level expansion is $\vert  \widetilde{R}\rangle=\vert\Delta\rangle+\cdots$ in this normalization.
The spherical four-point conformal block is then 
$\mathcal{B}[C_{0,4}]=\langle  \widetilde{R}(\Delta_4,\Delta_3;1)\vert  \widetilde{R}(\Delta_1,\Delta_2;z) \rangle$.
Two fundamental hypermultiplets therefore are associated with the regular state $\vert \widetilde{R}\rangle$.
The mass parameters of these matters are related with the Liouville momenta $\alpha_i$ 
of the corresponding primary states with conformal dimension $\Delta_i=\alpha_i(Q-\alpha_i)$ 
by\footnote{Here we adopt the standard convention of the Liouville momentum.
In the next subsection we {will change} 
the definition with $\Delta_i=\alpha_i(\alpha_i-Q)$ 
{for convenience}
.}
\begin{align}
&m_1=\alpha_1-\alpha_2-\frac{Q}{2},\quad
m_2=\alpha_1+\alpha_2-\frac{Q}{2},\quad\nonumber\\
&\tilde{m}_1=\alpha_3-\alpha_4-\frac{Q}{2},\quad
\tilde{m}_2=\alpha_3+\alpha_4-\frac{Q}{2},
\end{align}
where $Q$ is related to the central charge by $c= 1 - 6 Q^2$.

To describe a state corresponding to a single hypermultiplet,
let us decouple the matter with mass parameter $m_1$
by sending $m_{1} \to \infty$.
In addition, we have to fix the low-energy dynamical scale finite 
 in order to keep low-energy gauge theory dynamics.
Since the AGT dictionary for the UV gauge coupling constant $\tau_{\textrm{UV}}$
translates the moduli into $z=e^{2\pi i \tau_{\textrm{UV}}}$,
the dynamical scale below the energy scale $m_1$ is the dimensional transmutation parameter $zm_1 \equiv \Lambda$.
We therefore have to send $z$ to zero with this dynamical scale fixed.
We can translate this limit in the language of two-dimensional CFT as
\begin{align}
&\alpha_1-\alpha_2\to\infty,\quad z\to0,\quad
c_0=\alpha_1+\alpha_2,\quad c_1=(\alpha_1-\alpha_2)\,z
\end{align}
for certain fixed values $c_{\,0,1}$.
This decoupling procedure makes the two primary fields $V_{\,1,2}$ colliding 
and their momenta infinitely massive.

This collision limit simplifies the regular state $\vert  \widetilde{R}\rangle$
and the resulting conformal block $\langle  \widetilde{R}\vert  \widetilde{R} \rangle$.
Indeed the limit leads to the following state in the Verma module \cite{Marshakov:2009gn}:
\begin{align}
\vert  \widetilde{R}\rangle\to
\vert  I_1(m_2,\Lambda)\rangle=
\sum_{Y,n,p}
(2m_2+Q)^{n-2p}\left(\frac{\Lambda}{2}\right)^n
Q
\left(\Delta\right)
^{-1}_{[1^{n-2p}\cdot2^p],Y}
\vert \Delta,Y \rangle.
\label{MMM}
\end{align}
This state $\vert I_{N_f=1}\rangle$ describes a puncture associated with one hypermultiplet.
For instance, the scalar product $\langle \widetilde{R}(\Delta_4,\Delta_3)\vert I_{N_f=1} \rangle$ gives
the Nekrasov partition function for $SU(2)$ SQCD with $2+1$ flavors $\CW(A_{1}, C_{0,2,\{ 2 \}})$,
and $\langle I_{N_f=1} \vert I_{N_f=1}\rangle$ provides that with $1+1$ flavors $\CW(A_{1}, C_{0,0,\{ 2,2 \}})$.

The irregular state with {\it no hypermultiplet} $\vert I_{N_f =0} \rangle$ can also be obtained 
by a similar decoupling limit. 
By using these states we can formulate the correspondence 
for $SU(2)$ gauge theory with $N_{f} (\leq 3)$ hypermultiplets.
This is an extended version of the AGT correspondence to asymptotically free gauge theories, 
and has been proven in \cite{Hadasz:2010xp} for the $N_{f}=0,1,2$ cases.

In spite of the complexity of the expression \eqref{MMM},
the following simple conditions characterize the state $\vert  I_1(m,\Lambda)\rangle$:
\begin{align}
L_1\,\vert  I_1(m,\Lambda)\rangle=\left(m+\frac{Q}{2}\right)\Lambda\,\vert  I_1(m,\Lambda)\rangle,\quad
L_2\,\vert  I_1(m,\Lambda)\rangle=\Lambda^2\,\vert  I_1(m,\Lambda)\rangle.
\label{coherent}
\end{align}
Actually we can show  that \eqref{MMM} is the unique solution to the conditions \eqref{coherent} up to an overall factor.
To check the relation with the gauge theory quickly, one can see that the insertion of the energy-momentum tensor $T(z)$
into the conformal block is identical to the $\phi_{2}$ in the $\epsilon_{1,2} \to 0$ limit \cite{Alday:2009aq}.
Let us define the insertion of $T(z)$ into the irregular conformal block which we are considering as
  \bea
  \phi_{2}^{{\rm CFT}}(z)
   =     \lim_{\epsilon_{1,2} \to 0} \left< T(z) \right>,
  \eea
up to an irrelevant coefficient.
The above conditions \eqref{coherent} for $\vert I_{N_f=1} \rangle$ agree 
with the behavior of the irregular puncture of degree $2$ \eqref{Nf=1}. 
We can also check the agreement of the coherent state condition on the irregular state $\vert I_{N_f =0}\rangle$ 
with the puncture of degree $3/2$.

\subsection{Irregular states from the collision of primaries}

In this subsection we review the approach by Gaiotto-Teschner \cite{Gaiotto:2012sf}
to obtain the irregular states from the collision of primaries. 
Let us consider the state that is obtained by 
acting $n$ primaries (vertex operators) $V_{\Delta_i}(z_i)$ on the primary state;
\beq
\vert R_n \rangle := \prod_{i=1}^n V_{\Delta_i} (z_i) \vert \Delta_{n+1} \rangle.
\eeq
By acting the Virasoro generators on this state, we obtain
\beq
T_{+}(y) \vert{R_n} \rangle =
\left[ \sum_{i=1}^n\frac{\Delta_i}{(y-z_i)^2}+\frac{\Delta_{n+1}}{y^2}
+  \sum_{i=1}^n \frac{z_i}{y(y - z_i)} \frac{\partial}{\partial z_i} + \frac{L_{-1}}{y} \right]  \vert{R_n} \rangle.
\eeq
We study the behavior of this equation in the collision limit in order to show a characteristic
of the collision-induced irregular vector.

We will take a singular behavior of the state from the above expression 
and evaluate the limit-value of it.
For this purpose, we introduce
\beq\label{phi_sing}
\partial_y \phi_{\rm sing} := \sum_{i=1}^n \frac{\alpha_i}{y-z_i} + \frac{\alpha_{n+1}}{y},
\eeq
and
\beq
T_{\rm sing} (y) := (\partial_y \phi_{\rm sing} )^2 + Q~\partial^2_y \phi_{\rm sing},
\eeq
following \cite{Gaiotto:2012sf}.
Here we employ the convention $\Delta_i=\alpha_i(\alpha_i-Q)$ of \cite{Mironov:2009by}.
A redefinition of the state by
\beq
 \vert R_n \rangle = \prod_{i=1}^n z_i^{2 \alpha_i \alpha_{n+1}} \prod_{1 \leq i < j \leq n}
 (z_i - z_j)^{2 \alpha_i \alpha_j} \vert \widetilde{R_n} \rangle,
\eeq
simplifies the action of the \lq\lq positive\rq\rq\  part of the energy momentum tensor 
$\displaystyle{T_{+} (y) = \sum_{k \geq -1} y^{-2-k} L_k}$:
\beq
T_{+}(y) \vert \widetilde{R_n} \rangle =
\left[ T_{\rm sing}(y) +  \sum_{i=1}^n \frac{z_i}{y(y - z_i)} \frac{\partial}{\partial z_i} + \frac{L_{-1}}{y} \right]  \vert \widetilde{R_n} \rangle.
\eeq
Now we have
\beq
\partial_y \phi_{\rm sing} = \frac {P_n(y)} {y \prod_{i=1}^n (y - z_i)},
\eeq
where $P_n(y) := c_0~y^n + c_1 y^{n-1} + \cdots + c_n$ is a polynomial of $n$-th order in $y$ 
and the coefficients are given by
\beqa\label{c_k}
c_0 &=& \alpha_1  + \cdots + \alpha_{n} + \alpha_{n+1}, \nt
c_k &=& (-1)^k \sum_{1 \leq i_1 < \cdots < i_k \leq n} z_{i_1} \cdots z_{i_k}
\left( \sum_{j \notin \{ i_1, \ldots, i_k \}}
\alpha_j  \right), \qquad (1 \leq k \leq n).
 \eeqa
Note that $c_k$ is $k$-th order in $z_i$'s and linear in $\alpha_j$'s. 
We will take the limit $z_i \to 0$ and $\alpha_j \to \infty$, while keeping
$c_0, c_1, \ldots, c_n$ finite. Thus all the primaries are colliding at the origin 
and all the \lq\lq momenta\rq\rq\ becomes large, keeping the total momentum finite. 
Let us look at the Virasoro conditions on the limit state $ \vert \widetilde{R_n} \rangle 
\to \vert I_n (\alpha, c_i) \rangle$. The limit of $T_{\rm sing}$ is simply
\beq
T_{\rm sing} (y) \to \frac{1}{y^2} \left( \frac{c_n}{y^{n}} + \cdots + \frac{c_1}{y} + c_0 \right)^2 
- \frac{Q}{y^2} \left( \frac{(n+1)c_n}{y^{n}} + \cdots + \frac{2c_1}{y} + c_0 \right).
\eeq
The limit of the derivative terms is more involved. We use
\beq
\frac{z_i}{y(y-z_i)} \frac{\partial}{\partial z_i} = \sum_{j=1}^n \frac{z_i}{y(y-z_i)} \frac{\partial c_j}{\partial z_i}
\frac{\partial}{\partial c_j}.
\eeq
By evaluating the Euler derivative of $c_j$ which is at most the first order in each $z_i$, we find
\beq
\sum_{i=1}^n  \frac{z_i}{y(y-z_i)} \frac{\partial c_j}{\partial z_i} \frac{\partial}{\partial c_j}
=  \sum_{i=1}^n \frac{c_j^{(i)}}{y(y-z_i)}  \frac{\partial}{\partial c_j}, \label{derivative}
\eeq
where $c_j^{(i)}$ is the part of $c_j$ that contains $z_i$, or more explicitly
\beq
c_j^{(i)} = (-1)^j z_i \sum_{i_k \neq i, 1 \leq i_1 < \cdots < i_{j-1} \leq n} z_{i_1} \cdots z_{i_{j-1}}
\left( \sum_{ \ell \notin \{ i, i_1, \ldots, i_{j-1} \}} \alpha_\ell \right). 
\eeq
Since $c_j$ is of order $j$ in $z_i$'s, the Euler derivatives  give
the overall factor $j$. When we reduce \eqref{derivative}, the common denominator is 
$y(y-z_1) \cdots (y-z_n)$ and it is easy see that the leading term is $j c_j/y^2$.
The remaining terms also produce the higher $c_k,~k > j$ by discarding some of $\alpha_j$'s,
which vanish in the limit. Note that it has an additional power of $y$ whose degree is 
determined by the discrepancy of the order in $z_i$'s between $c_k$ and $c_j$.
Thus we see
\beq
\sum_{i=1}^n \frac{c_j^{(i)}}{y(y-z_i)}
\to \frac{j}{y^2} \left(c_j + \cdots + y^{j-n} c_n \right).
\eeq
In summary the limiting state satisfies
\beqa
T_{+} (y) \vert I_n \rangle
&=& \left[ ~\frac{1}{y^2} \left( \frac{c_n}{y^{n}} + \cdots + \frac{c_1}{y} + c_0 \right)^2 
- \frac{Q}{y^2} \left( \frac{(n+1)c_n}{y^{n}} + \cdots + \frac{2c_1}{y} + c_0 \right) \right. \CR
&&~~~ \left. + \sum_{j=1}^n \frac{j}{y^2} \left(c_j + \cdots + \frac{c_n}{y^{n-j}} \right)\frac{\partial}{\partial c_j}
+ \frac{L_{-1}}{y} \right] \vert I_n \rangle.
\eeqa
Looking at the coefficient of $y^{-2-k}$ we obtain the action of $L_k$ on $\vert I_n \rangle$ as follows;
\beqa
L_0 \vert I_n \rangle &=& \left[ c_0( c_0-Q) + \sum_{j=1}^n j c_j \frac{\partial}{\partial c_j} \right]  \vert I_n \rangle, \nt
L_k  \vert I_n \rangle &=& \left[ c_k (2 c_0-(k+1) Q ) 
+ \sum_{\ell=1}^{k-1} c_\ell c_{k-\ell} + \sum_{\ell =1}^{n-k}
\ell c_{\ell+k} \frac{\partial}{\partial c_{\ell}} \right] \vert I_n \rangle,  \quad ( 1 \leq k \leq n -1) \nt
L_{n}  \vert I_n \rangle &=& \left[ c_n (2 c_0-(n+1) Q) 
+ \sum_{\ell=1}^{n-1} c_\ell c_{n-\ell}  \right] \vert I_n \rangle,  \nt
L_{n+k}  \vert I_n \rangle &=& \left[  \sum_{\ell=k}^{n} c_\ell c_{n+ k-\ell}  \right] \vert I_n \rangle,\quad ( 1 \leq k \leq n) 
\label{VirasoroIrreg}
\eeqa
and $L_k  \vert I_n \rangle =0$ for $k > 2n$. 
We obtain an irregular state of order $n$ introduced by \cite{Gaiotto:2012sf}. 
We see that $\vert I_n \rangle$ is an eigenstate of $L_n, \ldots, L_{2n}$, but not for $L_0, \ldots, L_{n-1}$.

As we summarized in Appendix \ref{sec:U(1)}, the irregular state of the $U(1)$ current algebra is 
nothing but a familiar coherent state in the Fock  space of free boson. Hence, if we employ a free
field realization of the Virasoro algebra
\beq
L_n := \sum_{k \in \mathbb{Z}} :a_k a_{n-k}:  -  Q (n+1) a_n,
\eeq
in terms of the $U(1)$ current $J(z) = \displaystyle{\sum_{n \in \mathbb{Z}}} a_n z^{-n-1}$, 
we find a solution to the conditions \eqref{VirasoroIrreg} as a coherent state
\beq
a_k \vert I_n \rangle_{F} = c_i \vert I_n \rangle_{F}~~(1 \leq k \leq n), \qquad
a_\ell \vert I_n \rangle_{F} = 0~~(\ell \geq n),
\eeq
in the Fock space. Here we identify some of the creation operators $a_{-k}$
with the differential operator $k \frac{\partial}{\partial c_k}$, which affects
the prescription of the normal ordering. To make use of such a free field solution
to the irregular state for the construction of the irregular conformal block,
we have to understand 
the role of the screening operators \cite{Gaiotto:2012sf}.
For Virasoro regular states the treatment of the screening operators 
{in the matrix model was worked out in \cite{Dijkgraaf:2009pc, Mironov:2010ym, Itoyama:2010ki},}
and one can recast a Virasoro conformal block into a Dotsenko-Fateev integral.
This idea should work also for irregular blocks in the collision limit \cite{Nishinaka:2012kn}.
In the $\cW_3$ case, however, {a similar} 
handling of the screening operators is an open problem at the moment.


Note that $\vert I_1\rangle $ is nothing but the irregular state $\left| I_{N_{f}=1} \right>$ discussed in section 2.1. 
In fact  $\vert I_1\rangle $ satisfies
\beqa
L_0 \vert I_1 \rangle &=& \left[ c_0 ( c_0-Q) + c_1 \frac{\partial}{\partial c_1} \right] \vert I_1 \rangle, \nt
L_1 \vert I_1 \rangle &=& 2(c_0-Q) c_1 \vert I_1 \rangle, \qquad L_2 \vert I_1 \rangle = c_1^2  \vert I_1 \rangle.
\eeqa
The last two conditions should be compared with the condition \eqref{coherent}.
We recover the famous dictionary $m \sim c_0, Q \sim \epsilon_{+}/2$. 
Moreover we find $c_1 \sim \Lambda$. 
The identification $c_1 \sim \Lambda$ implies 
the first relation for $L_0$ can be regarded as the Matone's relation \cite{Matone:1995rx} in Seiberg-Witten theory.

The higher irregular states $\left| I_{n} \right>$ ($n \geq 2$) are argued to correspond to isolated SCFT's
$\CW(A_{1}, C_{0,1,\{ n+1 \} })$. 
Indeed if we put this irregular state at $z=0$, $\phi_{2}^{{\rm CFT}}$ behaves locally 
  \bea
 \lim_{\epsilon_{1,2} \to 0} \left< T(z) \right> =  \phi_{2}^{{\rm CFT}} (z)
   =     \frac{const}{z^{2n+2}} + \cdots,
  \eea
and this agrees with the behavior of $\phi_{2}$ of $\CW(A_{1}, C_{0,1,\{ n+1 \} })$ theory,
as we will see in section \ref{sec:SU(2)}.
The simplest state corresponding to the SCFT is $\left| I_{2} \right>$, 
whose conditions 
are explicitly given by
\beqa
L_0 \vert I_2  \rangle &=&  \left[ c_0 (c_0-Q) + c_1 \frac{\partial}{\partial c_1} 
+ 2 c_2 \frac{\partial}{\partial c_2}  \right] \vert I_1 \rangle, \nt
L_1 \vert I_2 \rangle &=& \left[ 2 c_1(c_0-Q) + c_2 \frac{\partial}{\partial c_1} \right] \vert I_1 \rangle, \nt
L_2 \vert I_2 \rangle &=& ( c_2 (2c_0-3Q) + c_1^2 ) \vert I_1 \rangle, \nt
L_3 \vert I_2 \rangle &=&  2 c_1 c_2  \vert I_2 \rangle,
\qquad
L_4 \vert I_2 \rangle = c_2^2  \vert I_2 \rangle.
\eeqa
Note that the eigenvalues of $L_{3,4}$ follow from the commutation relations
$[ L_2, L_1] \sim L_3$ and $[L_3, L_1] \sim 2 L_4$.

In \cite{Bonelli:2011aa} the explicit expressions like \eqref{MMM} for the higher irregular states have been found.
These expressions satisfy a set of similar  conditions described above.
(The conditions derived in \cite{Bonelli:2011aa} are slightly different from those here
because of a difference of  conventions, as we will explain in Appendix \ref{sec:BMTGT}.)
 In fact the irregular states in \cite{Bonelli:2011aa} are slightly different 
from those constructed by the collision limit here. 
This can be seen from the fact that the coefficient of the primary state in the expansion 
of the irregular state $\left| I_{n} \right>$ is a nontrivial function of the parameters $c_i$. 
On the other hand, in \cite{Bonelli:2011aa}, this was normalized to be 1.
We will discuss this point more in section \ref{sec:conclusions}.
The states in Virasoro module satisfying these conditions have also considered in \cite{Felinska:2011tn}.


\section{Irregular states of $\CW_3$ algebra}
\label{sec:W3}

\begin{figure}[t]
 \begin{center}
 \includegraphics[width=4cm,clip]{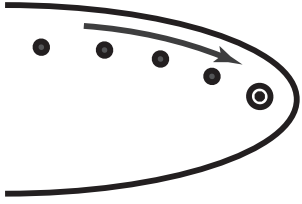}
 \end{center}
 \caption{The collision of vertex operators $V_1(z_1),\cdotsb,V_{n+1}(z_{n+1})$.}
 \label{fig:collision}
\end{figure}

In this section we generalize the story for the Virasoro algebra 
to irregular states in $\CW_3$ algebra and $SU(3)$ gauge theories. 
In this case there are two types of regular punctures~\cite{Gaiotto:2009we}.
The first one is of simple type and associated with a $U(1)$ flavor symmetry
(or carries a single mass parameter). The other is of full type
and has an $SU(3)$ flavor symmetry with two mass parameters. 
We are going to consider the collision of one full puncture
with $n$ simple punctures, as depicted in fig.~\ref{fig:collision}. 

Based on the $\CW_3$ Ward identities for
the primary states, we first show that the irregular state
obtained by a collision of a full puncture and a simple puncture
is nothing but the generalized Whittaker state introduced in \cite{Kanno:2012xt}. 
It is known that due to the special 
condition on the $A_2$ Toda momentum for the simple puncture~\cite{Wyllard:2009hg},
the primary state of the simple puncture has a level one null state,
which allows us to express the action of the mode $W_{-1}$ 
in terms of the differential operator in the coordinates of the punctures. 
Then we consider the case where $n$ simple punctures are colliding with a full puncture
and derive the conditions which should be satisfied by the irregular
states. For the Virasoro part the condition is the same as the $SU(2)$ 
case described above, since the corresponding Ward identities remain the same.
For the $\CW_3$ part, the condition involves the generators up to $W_{3n}$.
The irregular state is an eigenstate for $W_{2n}, \ldots, W_{3n}$
and the actions of $W_{n}, \ldots, W_{2n-1}$  are given by the first order 
differential operators in $c_0, c_1, \ldots, c_n$. Unfortunately
it seems that we cannot write the actions of the lower non-negative modes
$W_{0}, W_{1}, \ldots, W_{n-1}$ in a simple way. 
We will see that the irregular $\CW_{3}$ states considered in this section correspond 
to the $\CW(A_{2}, C_{0,1,\{ n+1 \}})$ theory in section \ref{sec:SU(3)}. 


\subsection{Collision of two punctures: $n=1$}

Let us start with the simplest example: the collision of two primary operators.
The state associated with two primary fields is
\begin{align}
\vert R_1 \rangle := V_{\vec{\alpha}_1} (z) \vert V_{\vec{\alpha}_2} (0) \rangle.
\end{align}
Since we will compute the scaling limit of the state
with keeping application to irregular conformal blocks in mind,
we work with chiral vertex operators $V_{\vec{\alpha}}(z)$,
instead of full primary fields.
To apply the AGT correspondence of $SU(3)$ gauge theories with fundamental hypermultiplets
\cite{Wyllard:2009hg},
we choose the operator $V_1$ semi-degenerate, {\em i.e.}
\begin{align}
&\Delta_{\vec{\alpha_i}} = \alpha_i^2 + \beta_i^2-Q^2, \qquad
w_{\vec{\alpha_i}} = \sqrt{\kappa}\alpha_i (\alpha_i^2  - 3 \beta_i^2),
\qquad \vec{\alpha}_1=\left(\alpha_1,-\frac{Q}{2}\right),
\end{align}
where $\kappa$ is given by the central charge $c=2-24\,Q^2$ as
\begin{align}
\kappa=\frac{32}{22+5c}.
\end{align}
The free field computation $Q=0$ therefore means $\kappa=1$.

Our collision limit is the same as the one in section \ref{sec:Virasoro}:
\begin{align}
\alpha_1\to\infty,\quad
z\to 0,\quad
c_0:=\alpha_1+\alpha_2,\quad
c_1:=\alpha_1z.
\end{align}
This limit for W-algebra
 is actually the same as the decoupling limit of one flavor which was studied in \cite{Kanno:2012xt}.
 Let us check it explicitly.
 The parameter identification of the AGT correspondence for $SU(3)$ SQCD is
\begin{align}
&m_1= -\frac{1}{\sqrt{3}}\alpha_1+\frac{Q}{2}+\frac{2}{\sqrt{3}}\alpha_2,\label{m1}\\
&m_2= -\frac{1}{\sqrt{3}}\alpha_1+\frac{Q}{2}-\frac{1}{\sqrt{3}}\alpha_2-\beta_2 , \\
&m_3= -\frac{1}{\sqrt{3}}\alpha_1+\frac{Q}{2}-\frac{1}{\sqrt{3}}\alpha_2+\beta_2,\label{m2}
\end{align}
where $m_{1,2,3}$ are three mass parameters for the fundamental hypermultiplets of the gauge theory.
With these, 
we can translate our collision limit in terms of the gauge theory parameters:
\begin{align}
&m_1\sim -\sqrt{3}\alpha_1\to -\infty,\\
&m_2= -\frac{c_0}{\sqrt{3}}+\frac{Q}{2}-\beta_2 ,\\
&m_3= -\frac{c_0}{\sqrt{3}}+\frac{Q}{2}+\beta_2,
\end{align}
with keeping $\Lambda\propto\alpha_1z$ finite.
This is precisely the decoupling limit of single hypermultiplet of $SU(3)$ SQCD.

Let us study the scaling limit of the state $\vert R_1 \rangle 
=:z^{ \Delta_{\textrm{int}}-\Delta_{\textrm{R}}+2\alpha_1\alpha_2+\frac{3Q^2}{4}}\vert \widetilde{R_1} \rangle$, 
where $\Delta_{\textrm{int}}=\alpha^2+\beta^2-Q^2$ and $\Delta_{\textrm{R}}=c_0^2+\beta_2^2-Q^2$.
Here we scale out the overall factor $z^{2\alpha_1\alpha_2+\frac{3Q^2}{4}}$
to get rid of diverging contribution.
The factor arises because
the contribution of
the state $\vert R_1 \rangle = V_{\vec{\alpha}_1} (z) \vert V_{\vec{\alpha}_2} (0) \rangle$ in a conformal block
with the internal momentum $ \Delta_{\textrm{int}}$ behaves as
\begin{align}
\vert R_1 \rangle \sim
z^{ \Delta_{\textrm{int}}- \Delta_1- \Delta_2}\left(\,\vert \Delta_{\textrm{int}} \rangle +\mathcal{O}(z) \,\right),
\end{align}
in other words, we work with the chiral vertex operator 
$V_{\vec{\alpha}_1}\,:\,\mathcal{V}_{\Delta_2} \to\mathcal{V}_{\Delta_{\textrm{int}}} $,
where $\mathcal{V}$ is the Verma module.
This means we expand the state $\vert R_1 \rangle $
 in the Verma module for the highest weight state $\vert \Delta_{\textrm{int}} \rangle$.
The power of the overall factor is then 
\begin{align}
 \Delta_{\textrm{int}}- \Delta_1- \Delta_2
 = \left(\Delta_{\textrm{int}}-\Delta_{\textrm{R}}\right)+2\alpha_1\alpha_2+\frac{3Q^2}{4},
\end{align}
and this factor gives a diverging contribution $2\alpha_1\alpha_2$ in the collision limit $\alpha_i\to\infty$.
Thus we scale it out by defining the renormalized state $\vert \widetilde{R_1} \rangle$.
The finite part $c_1^{ \Delta_{\textrm{int}}-\Delta_{\textrm{R}}}$
gives the classical contribution to the 
corresponding instanton partition function.
This normalization 
$\vert \widetilde{R_1} \rangle=\vert \Delta_{\textrm{int}} \rangle +\vert \textrm{descendants}\rangle$
has been used in the context of the Whittaker-Gaiotto states for asymptotically-free gauge theories 
\cite{Gaiotto:2009ma, Taki:2009zd, Keller:2011ek, Kanno:2012xt}.

Let us introduce the currents with \lq\lq positive\rq\rq\ modes
\begin{align}\label{def-TW+}
&T_{+} (y) := \displaystyle{\sum_{k \geq -1} y^{-2-k} L_k},\qquad
W_{+} (y) := \displaystyle{\sum_{k \geq -2} y^{-3-k} W_{k}}.
\end{align}
Then the action of the currents on the state leads to
\begin{align}
\label{T+}
&T_{+} (y) \vert R_1 \rangle = \left(\,
\frac{\Delta_1}{(y-z)^2} + \frac{\Delta_2}{y^2} 
+ \frac{z}{y(y-z)}\frac{\partial}{\partial z}  +   \frac{L_{-1}}{y} \,\right) \vert R_1 \rangle,\nonumber\\
&W_{+} (y) \vert R_1 \rangle = \left(\,
\frac{w_1}{(y-z)^3} + \frac{w_2}{y^3} 
+ \frac{W_{-1}^{(1)}}{(y-z)^2}  +   \frac{W_{-1}^{(2)}}{y^2}  
+ \frac{W_{-2}^{(1)}}{y-z}  +   \frac{W_{-2}^{(2)}}{y}  \,   \right) \vert R_1 \rangle.
\end{align}
This formula follows from OPE's between the current and the primary fields.
Here we used the fact that the action of $L_{-1}$ on a primary operator is just the differential $\partial_z$.
$W^{(i)}_{-k}$ is the generator acting only on the primary field $V_{\vec{\alpha}_i}$.
We rewrite the right hand side of the second equation by using the formulas
\begin{align}
&W_0 \vert R_1 \rangle 
=\left( w_1+w_2+2zW_{-1}^{(1)}+z^2W_{-2}^{(1)} \right)\vert R_1 \rangle ,\nt
&W_{-1} \vert R_1 \rangle 
=\left( W_{-1}^{(1)}+zW_{-2}^{(1)}+W_{-1}^{(2)} \right)\vert R_1 \rangle ,\nt
&W_{-2} \vert R_1 \rangle 
=\left( W_{-2}^{(1)}+W_{-2}^{(2)} \right)\vert R_1 \rangle.
\end{align}
Then we obtain
\begin{align}
W_{+} (y) \vert R_1 \rangle = \bigg(\,\frac{w_1}{(y-z)^3} + \frac{w_2}{y^3} 
+ 
\frac{z^2}{y^2(y-z)^2}&W_{-1}^{(1)} - \frac{w_1+w_2}{y^2(y-z)} \nonumber\\
&\label{W+}+ 
\frac{W_0}{y^2(y-z)} 
+   \frac{W_{-1}}{y^2}  
+   \frac{W_{-2}}{y}    \,\bigg) \vert R_1 \rangle.
\end{align}
At first sight, the right hand sides of (\ref{T+}) and (\ref{W+}) seem to diverge in the collision limit.
To evaluate the limit values correctly,
we introduce the following combinations
\begin{align}
&T_{\textrm{sing}}(y):=\left(\partial \phi_1(y)\right)^2+\left(\partial \phi_2(y)\right)^2,\nt
&W_{\textrm{sing}}(y):=\sqrt{\kappa}\left((\partial \phi_1(y))^3
-3\partial \phi_1(y)(\partial \phi_2(y))^2\right),
\end{align}
where 
\begin{align}
\partial \phi_1(y)=\frac{\alpha_1}{y-z}+\frac{\alpha_2}{y}
,\quad
\partial \phi_2(y)=\frac{\beta_1}{y-z}+\frac{\beta_2}{y},
\end{align}
and $\beta_1=-Q/2$.
The point is that these combinations remain finite in the collision limit.
Let us start with computing the contribution of the stress-energy current $T_+(y)\vert R_1\rangle$.
By using $T_{\textrm{sing}}(y)$ we can recast it into
\begin{align}
T_+(y)\vert R_1\rangle&=
\left(\,T_{\textrm{sing}}(y)+   \frac{L_{-1}}{y}
-\frac{Q^2}{(y-z)^2} - \frac{Q^2}{y^2} 
-2\frac{\alpha_1\alpha_2+\beta_1\beta_2}{y(y-z)}
+ \frac{z}{y(y-z)}\frac{\partial}{\partial z}  \right)\vert R_1 \rangle,\nonumber\\
&=z^{\Delta_{\textrm{int}}-\Delta_{\textrm{R}}+2\alpha_1\alpha_2+\frac{3Q^2}{4}}
\bigg(\,T_{\textrm{sing}}(y)+   \frac{L_{-1}}{y}
-\frac{Q^2}{(y-z)^2} - \frac{Q^2}{y^2} \nonumber\\
&\qquad\qquad\qquad\qquad\quad\,\,\,\,
-\frac{2\beta_1\beta_2-\frac{3Q^2}{4}-\Delta_{\textrm{int}}+\Delta_{\textrm{R}}}{y(y-z)}
+ \frac{z}{y(y-z)}\frac{\partial}{\partial z}  \,\bigg)
\vert \widetilde{R_1} \rangle.
\end{align}
Notice that due to the re-normalization of the state,
the diverging term is completely canceled out from the above expression.
Since $\partial\phi_1\to c_1/y^2+c_0/y$ and $\partial\phi_2\to (\beta_2-Q/2)/y$ in the collision limit,
we obtain the following limit value for $T_+$:
\begin{align}
T_+(y)\vert \widetilde{R_1} \rangle=
\left(\,
\frac{c_1^2}{y^4}+\frac{2c_0c_1}{y^3}
+ \frac{c_1\frac{\partial}{\partial c_1}+\Delta_{\textrm{int}} }{y^2}
+   \frac{L_{-1}}{y}\right)
\vert \widetilde{R_1} \rangle.
\end{align}
It is immediately obvious from this formula that
the irregular state $\vert \widetilde{R_1} \rangle$ is characterized by the following conditions:
\begin{align}
&L_2\vert \widetilde{R_1} \rangle=
c_1^2\vert \widetilde{R_1} \rangle, \label{L2} 
\\
&L_1\vert \widetilde{R_1} \rangle=
2c_0c_1\vert \widetilde{R_1} \rangle, \label{L1} 
\\
&L_0\vert \widetilde{R_1} \rangle=
\left(c_1\frac{\partial}{\partial c_1}+\Delta_{\textrm{int}}
\right)
\vert \widetilde{R_1} \rangle.
\end{align}
Since the AGT dictionary used in \cite{Wyllard:2009hg, Kanno:2012xt}
implies the parametrization $c_0=\frac{\sqrt{3}}{2}(Q-m_2-m_3)$ and $\beta_2=\frac{m_3-m_2}{2}$,
it is easy to check that these conditions are actually equal to those in \cite{Kanno:2012xt}.
The dynamical scale there is $\Lambda=-i\sqrt{3}c_1$.

Let us move on to computation of $W_{+}$ side.
With some algebra, we get
\begin{align}
W_{\textrm{sing}}(y)
=&\frac{w_1}{(y-z)^3}+\frac{w_2}{y^3}\nonumber\\
&+3\sqrt{\kappa}\left(
\frac{\alpha_2\left(\alpha_1^2-\frac{Q^2}{4}\right)+Q\alpha_1\beta_2}{(y-z)^2y}
+\frac{\alpha_1\left(\alpha_2^2-\beta_2^2\right)+Q\alpha_2\beta_2}{(y-z)y^2}
\right).
\end{align}
We can therefore use this formula to eliminate the terms $\frac{w_1}{(y-z)^3}+\frac{w_2}{y^3}$
from (\ref{W+}).
We can also rewrite $W_{-1}^{(1)}$  in the right hand side of (\ref{W+}) into a differential operator.
Since the generator $W_{-1}$ acts on the semi-degenerate field as the operator $L_{-1}$,
we obtain
\begin{align}
W_{-1}^{(1)}\vert R_1 \rangle
&=\frac{3w_1}{2\Delta_1}L_{-1}^{(1)}\vert R_1 \rangle
=\frac{3\sqrt{\kappa}\alpha_1}{2}\partial_z\vert R_1 \rangle\nonumber\\
&=z^{\Delta_{\textrm{int}}-\Delta_{\textrm{R}}+2\alpha_1\alpha_2+\frac{3Q^2}{4}-1}
\frac{3\sqrt{\kappa}}{2}\alpha_1
(2\alpha_1\alpha_2+\frac{3Q^2}{4}+\Delta_{\textrm{int}}-\Delta_{\textrm{R}}+z\partial_z)\vert \widetilde{R_1} \rangle.
\end{align}
By combining these results,
we can recast (\ref{W+}) into the following form
\begin{align}
W_{+} (y)\vert R_1 \rangle 
 = z^{\Delta_{\textrm{int}}-\Delta_{\textrm{R}}+2\alpha_1\alpha_2+\frac{3Q^2}{4}}\bigg( W_{\rm sing}(y) 
 +&\frac{3\sqrt{\kappa}z\alpha_1}{2y^2(y-z)^2} z\frac{\partial}{\partial z}
 +\frac{\sqrt{\kappa}(yP_1+P_2)}{y^2(y-z)^2} 
  \nonumber \\
&\qquad
+ 
\frac{W_0}{y^2(y-z)} 
+   \frac{W_{-1}}{y^2}  
+   \frac{W_{-2}}{y}
   \bigg) \vert \widetilde{R_1} \rangle,\nonumber
\end{align}
where
\begin{align}
&P_1=-c_0^3+3c_0\left(\beta_2^2-Q\beta_2+\frac{Q^2}{4}\right),\nt
&P_2=(c_0^3-3\beta_2^2c_0)z+3Q\beta_2(\alpha_2z)+\frac{3Q^2}{8}(\alpha_1z)
+\frac{3}{2}(\alpha_1z)(\Delta_{\textrm{int}}-\Delta_{\textrm{R}}).
\end{align}
Then, the collision limit of $W_{+} (y) \vert \widetilde{R_1} \rangle $ is obviously finite and the explicit form is
\begin{align}
 W_{+}(y)\vert \widetilde{R_1} \rangle 
 = &\bigg(\, W_{\rm sing}(y) 
 +\frac{\sqrt{\kappa}c_1}{y^4}
 \left(
-3Q\beta_2+\frac{3Q^2}{8}+\frac{3}{2}(\Delta_{\textrm{int}}-\Delta_{\textrm{R}})
+ \frac{3}{2}c_1\frac{\partial}{\partial c_1} \right)
\nonumber\\
&\label{W+2}+ 
\frac{W_0+\sqrt{\kappa}\left(-c_0^3+3c_0(\beta_2 -Q/2)^2\right) }{y^3} 
+   \frac{W_{-1}}{y^2}  
+   \frac{W_{-2}}{y}  
  \,\bigg) \vert \widetilde{R_1} \rangle.
\end{align}

Let us read off the condition for the irregular state from the formula (\ref{W+2}).
Since the collision limit leads to
$\partial\phi_1\to c_0 /y+c_1/y^2$ and $\partial\phi_2\to (\beta_2 -Q/2)/y$,
the limit value of $W_{\rm sing}(y)$ takes the following form
\begin{align}
W_{\rm sing}(y) 
=&\sqrt{\kappa}
\left(
\frac{c_1^3}{y^6}
+\frac{3c_0c_1^2}{y^5}
+\frac{c_1(3c_0^2
-3(\beta_2 -Q/2)^2)
}{y^4}
+\frac{c_0^3
-3(\beta_2 -Q/2)^2c_0
}{y^3}
\right).
\end{align}
This result implies
that the irregular state $\vert \widetilde{R_1} \rangle $ satisfies
\begin{align}
&  W_1\vert \widetilde{R_1} \rangle 
 =\frac{3\sqrt{\kappa}c_1}{2}\left(  c_1\frac{\partial}{\partial c_1}
 +c_0^2-3\beta_2^2+\frac{3Q^2}{4}+\Delta_{\textrm{int}}
 \right)\vert \widetilde{R_1} \rangle, \\
& W_2\vert \widetilde{R_1} \rangle 
 =3\sqrt{\kappa}c_0 c_1^2\vert \widetilde{R_1} \rangle,  \label{W2} \\
& W_3\vert \widetilde{R_1} \rangle 
 =\sqrt{\kappa}c_1^3\vert \widetilde{R_1} \rangle \label{W3}.
\end{align}
We can rewrite the first condition as
\begin{align}
W_1\vert \widetilde{R_1} \rangle 
 =\frac{3\sqrt{\kappa}c_1}{2}\left(  L_0
{ +c_0^2-3\beta_2^2+\frac{3}{4}Q^2}
\right)\vert \widetilde{R_1} \rangle \label{W1} .
\end{align}
Since 
$
c_0^ 2 - 3 \beta_2^2 + \frac{3}{4} Q^2 = \frac{3}{2} ( Q^2 -  Q (m_2 + m_3) + 2 m_2 m_3 )
$
from \eqref{m1}\,--\,\eqref{m2},
the conditions \eqref{W2}, \eqref{W3} and \eqref{W1} are also exactly the same as 
those for the generalized Whittaker state introduced in \cite{Kanno:2012xt} with $i \Lambda= \sqrt{3}c_1$:
\begin{align}
& 
 W_1\vert \widetilde{R_1} \rangle 
 =\frac{\sqrt{3\kappa}i\Lambda}{2}\left(  L_0
+\frac{3}{2}
\left(2m_2m_3-Q(m_2+m_3)+Q^2 \right)
\right)\vert \widetilde{R_1} \rangle,\nt
& W_2\vert \widetilde{R_1} \rangle 
 =\frac{\sqrt{3\kappa} (i\Lambda)^2}{2}(Q-m_2-m_3)\vert \widetilde{R_1} \rangle, \nt
& W_3\vert \widetilde{R_1} \rangle 
 =\frac{\sqrt{3\kappa}(i\Lambda)^3}{9}\vert \widetilde{R_1} \rangle.
\end{align}
%
%

It is easy to check that the five conditions \eqref{L2}, \eqref{L1}, \eqref{W2}, \eqref{W3} and \eqref{W1} 
for $\vert \widetilde{R_1} \rangle $ are consistent with the $\CW_3$ algebra\footnote{See Appendix A 
for our conventions of $\CW_3$ algebra.}.
Note that the $\CW_3$ algebra is generated by $L_{1,2}$ 
and $W_1$ by multiple commutators. Due to the presence of $L_0$ term in \eqref{W1} 
the commutation relation $[ L_{n-1}, W_1] = (2n-3) W_n$ implies the non-vanishing eigenvalues of $W_2, W_3$. 
Furthermore, one should have
\beqa
\left[ W_3, W_1 \right] &=& \frac{18}{4 -15 Q^2} (L_2)^2 =  \frac{18}{4 -15 Q^2} c_1^4, \nt
\left[ W_2, W_1 \right] &=& \frac{18}{4 -15 Q^2} L_1 L_2 =  \frac{36}{4 -15 Q^2} c_0 c_1^3.
\eeqa
These are consistent with
\beq
\left[ W_{2,3} , W_1 - \frac{3\sqrt{\kappa}c_1}{2} L_0 \right] =0.
\eeq


\subsection{Collision of three punctures: $n=2$}
 
We next compute the collision limit of two simple-type punctures and a single full-type one.
In view of the result in the Virasoro case \cite{Gaiotto:2012sf} we expect that
the irregular state from the collision of more than two punctures gives rise to an isolated SCFT.
In order to work out the 
correspondence with the isolated SCFT coming from the linear quiver theory to be 
discussed in section 5, we will derive the defining condition for the $\CW_3$ irregular state.
In the language of the two-dimensional Toda CFT,
these three punctures are described by the state
\begin{align}
&\vert R_2 \rangle := V_{\vec{\alpha}_1} (z_1)V_{\vec{\alpha}_2} (z_2)  \vert V_{\vec{\alpha}_3} (0) \rangle,\quad
\vec{\alpha}_{1,2}
=\left(\,\alpha_{1,2},-\frac{Q}{2}\,\right),
\end{align}
where $V_{\valp_{1,2}}$ are 
semi-degenerate fields associated with the simple punctures.
The collision limit of our interest is described by the following scaling limit:
\begin{align}
&\alpha_i\to\infty\,~\textrm{ for }~i=1,2,3,\quad
z_i\to 0\,~\textrm{ for }~i=1,2,3,\nt
&\textrm{with fixing}\quad
\alpha=\alpha_1+\alpha_2+\alpha_3,\quad
c_1=\alpha_1z_1+\alpha_2z_2,\quad
c_2=\alpha_3z_1z_2\quad\textrm{finite}.
\end{align}

Let us study the action of the W-current on the resulting irregular state
by computing the scaling limit of the corresponding state:
\begin{align}
W_{+} (y) \vert R_2 \rangle &= \Big[
{\frac{w_1}{(y-z_1)^3} +\frac{w_2}{(y-z_2)^3} + \frac{w_3}{y^3} }
\nonumber\\&\hspace{10mm}\label{W+3}
+ 
{\frac{W_{-1}^{(1)}}{(y-z_1)^2}+ \frac{W_{-1}^{(2)}}{(y-z_2)^2}   +   \frac{W_{-1}^{(3)}}{y^2}}
  +
{\frac{W_{-2}^{(1)}}{y-z_1}  +  \frac{W_{-2}^{(2)}}{y-z_2} +  \frac{W_{-2}^{(3)}}{y} }
   \Big] \vert R_2 \rangle.
\end{align}
Note that the computation of 
the action of the energy-momentum tensor $T_+(y)$ is completely parallel to that of Liouville theory
that was reviewed in section 2.2.
As we did in the case of $n=1$, we can rewrite the second line of the right hand side as
\ba
\textrm{
{Second Line}
}&=&
{
\frac{(2yz_1-z_1^2)W_{-1}^{(1)}}{y^2(y-z_1)^2}
+ \frac{(2yz_2-z_2^2)W_{-1}^{(2)}}{y^2(y-z_2)^2}
+   \frac{
W_{-1}
}{y^2}
}
+ 
{ \frac{W_{-2}}{y}}
\nonumber\\
&&+
{
\frac{z_1^2W_{-2}^{(1)}}{y^2(y-z_1)}
+ \frac{z_2^2W_{-2}^{(2)}}{y^2(y-z_2)}  
}
=: S(y).
\ea
In order to eliminate $W_{-2}^{(i)}$,
which does not act as an differential operator,
we use
\begin{align}
&W_0\vert{R}_2 \rangle 
=\left(w_1+w_2+w_3+2z_1W_{-1}^{(1)}+2z_2W_{-1}^{(2)}
+z_1^2W_{-2}^{(1)}+z_2^2W_{-2}^{(2)}\right)\vert{R}_2 \rangle,  \nonumber\\
&W_1\vert{R}_2 \rangle
=\left(3z_1w_1+3z_2w_2+3z_1^2W_{-1}^{(1)}+3z_2^2W_{-1}^{(2)}
+z_1^3W_{-2}^{(1)}+z_2^3W_{-2}^{(2)}\right)\vert{R}_2 \rangle .
\end{align}
Then we can rewrite the the above equation $S(y)$ as
\ba
S(y)&=&
\frac{W_{-2}}{y}+  \frac{W_{-1}}{y^2}
+\frac{(y-z_1-z_2)W_{0}}{y^2(y-z_1)(y-z_2)}
+\frac{W_{1}}{y^2(y-z_1)(y-z_2)}
\nonumber\\
&&{+\frac{(-y-2z_1+z_2)w_{1}+(-y+z_1-2z_2)w_{2}+(-y+z_1+z_2)w_{3}}{y^2(y-z_1)(y-z_2)}}
\nonumber\\\label{Sy}
&&{+
\frac{z_1^2(z_1-z_2)W_{-1}^{(1)}}{y^2(y-z_1)^2(y-z_2)}
+ \frac{z_2^2(z_2-z_1)W_{-1}^{(2)}}{y^2(y-z_1)(y-z_2)^2} 
}.
\ea
To get rid of the classical contribution
to take the scaling limit, we introduce  $ \vert \widetilde{R_2} \rangle$ as
\begin{align}\label{R2}
 \vert R_2 \rangle = z_1^{2 \alpha_1 \alpha_3} z_2^{2 \alpha_2 \alpha_3}
 (z_1 - z_2)^{2 \alpha_1\alpha_2} \vert \widetilde{R_2} \rangle.
\end{align}
This is the same as the Virasoro case \cite{Gaiotto:2012sf}.
We should mention that there exists an ambiguity in the choice of this prefactor.
This choice will affect the overall factor $C$ of the normalized state
and the resulting irregular state as follows:
\begin{align}
\vert \widetilde{R_2} \rangle
=C(c_i)\, \vert \Delta \rangle+\cdots.
\end{align}
The correct choice of the normalization must be fixed, for example, so that
the scalar products of the irregular states can reproduce the Nekrasov partition functions of 
the corresponding isolated SCFT's. However, 
it is not clear that what is {a correct definition} of the Nekrasov partition function
of such SCFT in general.
Hence, in the following we assume that the choice in \cite{Gaiotto:2012sf} works 
{also for $\cW_3$ case.} 

Let us move on to the computation of the limit value of the normalized 
state with the W-action $W_{+}\vert \widetilde{R_2} \rangle$.
By using the explicit action of generators $W_{-1}^{(1,2)}$, 
which is a differential operator on $ \vert{R_2} \rangle$,
we obtain the following expression for the last line of (\ref{Sy})
\begin{align}
&\left(
\frac{z_1^2(z_1-z_2)W_{-1}^{(1)}}{y^2(y-z_1)^2(y-z_2)}
+ \frac{z_2^2(z_2-z_1)W_{-1}^{(2)}}{y^2(y-z_1)(y-z_2)^2} 
\right)\vert{R}_2 \rangle\nonumber\\&
=
z_1^{2 \alpha_1 \alpha_3} z_2^{2 \alpha_2 \alpha_3}
 (z_1 - z_2)^{2 \alpha_1\alpha_2} 
 \,\Big(\,
 \frac{3z_1\alpha_1(z_1-z_2)z_1\partial_{z_1}}{2y^2(y-z_1)^2(y-z_2)}
 + \frac{3z_2\alpha_2(z_2-z_1)z_2\partial_{z_2}}{2y^2(y-z_1)(y-z_2)^2}\nonumber\\
& +
{
  \frac{3\alpha_1^2\alpha_3z_1(z_1-z_2)+3\alpha_1^2\alpha_2z_1^2}{y^2(y-z_1)^2(y-z_2)}
  +
  \frac{3\alpha_2^2\alpha_3z_2(z_2-z_1)+3\alpha_2^2\alpha_1z_2^2}{y^2(y-z_1)(y-z_2)^2}
  }
  \label{LastTwoLines}
 \Big)
 \vert \widetilde{R_2} \rangle.
\end{align}
In the scaling limit these terms become
\begin{align}
 \frac{3z_1\alpha_1(z_1-z_2)z_1\partial_{z_1}}{2y^2(y-z_1)^2(y-z_2)}
 + \frac{3z_2\alpha_2(z_2-z_1)z_2\partial_{z_2}}{2y^2(y-z_1)(y-z_2)^2}
 \to
\frac{ 3c_1c_2\partial_{c_1}+3c_2^2\partial_{c_2}}{y^5}
+{\frac{ 3c_2^2\partial_{c_1}}{2y^6}},
\end{align}
and we can also easily evaluate the limit values of the remaining terms in (\ref{Sy}) only with a little algebra.

To derive the limit value of the first line of (\ref{W+3}),
let us introduce $W_{\textrm{sing}}(y)$ for the case $n=2$ as follows:
\begin{align}
\partial \phi_1(y)=\frac{\alpha_1}{y-z_1}+\frac{\alpha_2}{y-z_2}+\frac{\alpha_3}{y}
,\quad
\partial \phi_2(y)=\frac{\beta_1}{y-z_1}+\frac{\beta_2}{y-z_2}+\frac{\beta_3}{y}.
\end{align}
The definition of  $W_{\textrm{sing}}$  in terms of $\partial \phi_i$ is precisely the same as the case of $n=1$.
Notice that since the primary fields $V_{1,2}$ are now semi-degenerate,
we set $\beta := \beta_{1}=\beta_2=-\frac{Q}{2}$.
With some algebra we can show
\ba
W_{\textrm{sing}}(y)
=\frac{w_1}{(y-z_1)^3}+\frac{w_2}{(y-z_2)^3}+\frac{w_3}{y^3}+\frac{
\left(\sum_{i=0}^3y^{3-i}\left(A^{(i)}(\alpha)+B^{(i)}(\alpha,\beta)\right)
\right)
}{y^2(y-z_1)^2(y-z_2)^2},
\ea
Using this equation, we can recast the first line of  (\ref{W+3}) in the function of $W_{\textrm{sing}}$, $A^{(i)}$ and $B^{(i)}$.
The coefficient polynomials $A^{(i)}$ and $B^{(i)}$
are given by
\begin{align}
&A^{(0)}+B^{(0)}=
\sqrt{\kappa}\alpha^3-w_1-w_2-w_3
-3\sqrt{\kappa}\alpha(\beta_3+2\beta)^2,\nonumber\\
&A^{(1)}=3\sqrt{\kappa}
\Big(
z_1
\left( 
-\alpha_1\alpha_2^2-\alpha_1\alpha_3^2
-2\alpha_2\alpha_3^2-2\alpha_2^2\alpha_3
-2\alpha_1\alpha_2\alpha_3
\right)\nonumber\\
&\qquad\qquad\,+z_2
\left( 
-\alpha_2\alpha_1^2-\alpha_2\alpha_3^2
-2\alpha_1\alpha_3^2-2\alpha_1^2\alpha_3
-2\alpha_1\alpha_2\alpha_3
\right)
\Big),\nonumber\\
&B^{(1)}=\sqrt{\kappa}z_1\Big(
3\alpha_1\beta^2-6\alpha_2\beta^2-6\alpha_3\beta_3^2
+6\alpha(\beta_3+\beta)(\beta_3+2\beta)-3z_1(\beta_3+2\beta_2)^2
\Big)+\Big(\,1\leftrightarrow2\,\Big),
\nonumber\\
&A^{(2)}=3\sqrt{\kappa}
\Big(z_1^2
\alpha_2\alpha_3(\alpha_2+\alpha_3)
+z_2^3\alpha_1\alpha_3(\alpha_1+\alpha_3)
+2z_1z_2\alpha_3\left(\alpha_1\alpha_2+\alpha_1\alpha_3+\alpha_2\alpha_3 \right)\Big),\nonumber\\
&B^{(2)}=\sqrt{\kappa}\Big(
z_1^2\left(
-6\alpha_1\beta^2
+3\alpha_2\beta^2+3\alpha_3\beta_3^2-6\alpha\beta\beta_3
+3\alpha_1(3\beta^2+2\beta\beta_3+\beta_3^2)\right)+\Big(\,1\leftrightarrow2\,\Big),\nonumber\\
&\qquad\qquad+z_1z_2
\left(
9\alpha_3\beta_3^2-\alpha_3(3\beta_3^2+18\beta\beta_3+6\beta^2)-6\alpha\beta_3(\beta_3+\beta)\right)\Big),
\nonumber\\
&
A^{(3)}=-3\sqrt{\kappa}z_1z_2\alpha_3^2\left(\alpha_1z_2+\alpha_2z_1 \right),\nonumber\\
\label{Wsing}
&B^{(3)}=\sqrt{\kappa}\Big(
z_1^2z_2\left(6\alpha_1\beta^2-3\alpha_2\beta^2-3\alpha_3\beta_3^2+
3\alpha_1(3\beta^2-\beta_3^2)+3\alpha(\beta_3^2-\beta^2)+3\alpha_3(2\beta\beta_3-\beta^2)\right)\nonumber\\
&\qquad\qquad+\Big(\,1\leftrightarrow2\,\Big)
+3z_1z_2(z_1\alpha_1+z_2\alpha_2)\left(3\beta^2-\beta_3^2 \right)\Big)
.
\end{align}
It is not so hard to take the scaling limit of these polynomials.
So the remaining task
is the evaluation of the limit of the term $W_{\textrm{sing}}(y)$.
The limit of $\partial\phi_{i}$ are easily evaluated as
\begin{align}
\partial\phi_{1}(y)=\frac{c_0}{y}+\frac{c_1}{y^2}+\frac{c_2}{y^3},\quad
\partial\phi_{2}(y)=\frac{\beta_3+2\beta}{y}
,
\end{align}
and by substituting them into the definition equation of $W_{\textrm{sing}}(y)$,
we can show that $W_{\textrm{sing}}$ takes the following form in the collision limit:
\begin{align}
&W_{\textrm{sing}}(y)=
\frac{\sqrt{\kappa}c_2^3}{y^9}
+\frac{3\sqrt{\kappa}c_1c_2^2}{y^8}
+\frac{3\sqrt{\kappa}c_2(c_2c_0+c_1^2)}{y^7}
+\frac{\sqrt{\kappa}(6c_0 c_1c_2+c_1^3)}{y^6}\nonumber\\
&\quad+\frac{3\sqrt{\kappa}\left(c_2c_0^2+c_1^2c_0-c_2(\beta_3+2\beta)^2\right)}{y^5}
+\frac{3\sqrt{\kappa}c_1\left(v^2-(\beta_3+2\beta)^2\right)}{y^4}
+\frac{\sqrt{\kappa}c_0(c_0^2-3(\beta_3+2\beta)^2)}{y^3}.
\end{align}

Now that we have the limit value of all terms of the right hand side of (\ref{W+3}),
we can rewrite down the explicit form of $W_{+}\vert \widetilde{R_2} \rangle$
in the collision limit.
We introduce $\vert I_2\rangle:=\lim_{\textrm{collision}}\vert \widetilde{R_2} \rangle$
to distinguish between before and after the limit. 
By substituting the above results into (\ref{W+3}),
we obtain the generating function of the irregular state condition for $ \vert I_2 \rangle$:
\begin{align}
W_{+}(y)\,
 \vert I_2 \rangle
 =
 \Big(
&\frac{W_{-2}}{y}+\frac{W_{-1}}{y^2}+\frac{W_{0}}{y^3}
+\frac{W_{1}}{y^4}\nonumber\\
&
+\sqrt{\kappa}\,\frac{3c_0^2c_2+3c_0 c_1^2-3(\beta_3^2+5\beta^2)c_2
+3c_1c_2\partial_{c_1}+3c_2^2\partial_{c_2}}{y^5}
\nonumber\\
&
+\sqrt{\kappa}\,\frac{6c_0 c_1c_2+c_1^3+
{3c_2^2\partial_{c_1}/2}}{y^6}
+\sqrt{\kappa}\,\frac{3c_0 c_2^2+3c_1^2c_2}{y^7}
+\sqrt{\kappa}\,\frac{3c_1c_2^2}{y^8}
+\sqrt{\kappa}\,\frac{c_2^3}{y^9}
 \Big) \vert I_2 \rangle.
\end{align}
This equation imposes the non-zero irregular state conditions on $\vert I_2 \rangle$
for the generators $W_{n=2,3,\ldots,6}$
\begin{align}
W_2
 \vert I_2 \rangle
 &=
 \sqrt{\kappa}\left({3c_0^2c_2+3c_0 c_1^2-3(\beta_3^2+5\beta^2)c_2
+3c_1c_2\partial_{c_1}+3c_2^2\partial_{c_2}}\right) \vert I_2 \rangle,
\nonumber\\
W_3
 \vert I_2 \rangle
&=\sqrt{\kappa}\,\left({6c_0 c_1c_2+c_1^3+
{3c_2^2\partial_{c_1}/2}} \right)\vert I_2 \rangle,\nonumber\\
W_4
 \vert I_2 \rangle
&=\sqrt{\kappa}\,\left({3c_0 c_2^2+3c_1^2c_2}\right)\vert I_2 \rangle,\nonumber\\
W_5
 \vert I_2 \rangle
&=\sqrt{\kappa}\,{3c_1c_2^2}\vert I_2 \rangle,\nonumber\\
W_6
 \vert I_2 \rangle
&=\sqrt{\kappa}\,{c_2^3}\vert I_2 \rangle.
\end{align}
Note that $\vert I_2 \rangle$ is
annihilated by the higher modes $W_{n >6}$. 
The conditions for the Virasoro generators $L_n$
are completely the same as those of the Liouville theory.
Once one fixes the ansatz for the irregular state
$\vert I_2\rangle=C(c_i)\vert\Delta\rangle+\cdots$,
we can use these irregular state conditions
to determine the irregular state explicitly.

 
\subsection{Collision of general $n+1$ punctures}

The computation in the previous subsections is generalized to the case of $n+1$ punctures. 
We consider here the collision limit of $n$ simple-type punctures and a full-type one.
In $A_2$ Toda CFT, the state with these punctures is defined as 
\ba
|R_n\rangle:=V_{\valp_1}(z_1)\cdots V_{\valp_n}(z_n)|V_{\valp_{n+1}}(0)\rangle,
\label{W3Rn}
\ea
where $V_{\valp_1},\ldots,V_{\valp_n}$ correspond to simple punctures
 and $V_{\valp_{n+1}}$ corresponds to a full puncture. 
This means that the momenta of these vertex operators satisfy 
$\valp_i=(\alpha_i,-Q/2)$ for $i=1,\ldots,n$ and $\valp_{n+1}=(\alpha_{n+1},\beta_{n+1})$. 
The collision limit of our interest is the following scaling limit:
\ba\label{limit}
\alpha_i\to\infty\,,\quad
z_i\to 0\,,
\ea
with their combinations given in (\ref{c_k}) kept finite:
\ba\label{cp}
c_p=(-1)^p\sum_{i=1}^{n+1}\,\Biggl(\alpha_i 
\sum_{\substack{1\leq j_1<\cdots<j_p\leq n\\[+1pt]j_1,\ldots,\,j_p\neq i}}
z_{j_1}\cdots z_{j_p}\Biggr).
\ea

The action of the ``positive'' W-current (\ref{def-TW+}) on the state \eqref{W3Rn} is 
\ba\label{W+Rn}
W_+(y) |R_n\rangle
=\sum_{j=1}^{n+1}\left(
 \frac{w_j}{(y-z_j)^3}+\frac{\Wj_{-1}}{(y-z_j)^2}+\frac{\Wj_{-2}}{y-z_j}
 \right)|R_n\rangle,
\ea
where we set $z_{n+1}=0$. Note that $V_{\valp_i}$ ($i=1,\ldots,n$) satisfies the degenerate condition
\ba\label{degenerate}
\Wi_{-1}|R_n\rangle=\frac{3w_{i}}{2\Delta_{i}}\Li_{-1}|R_n\rangle,
\ea
where the conformal weights are
\ba\label{weight}
&&\back
\Delta_i=\alpha_i^2-\frac34Q^2\,,\quad
w_i=\sqrt{\kappa}
 \alpha_i\left(\alpha_i^2-\frac34Q^2\right)\quad(\text{for}~i=1,\ldots,n),
\nt&&\back
\Delta_{n+1}=\alpha_{n+1}^2+\beta_{n+1}^2-Q^2\,,\quad
w_{n+1}=\sqrt{\kappa}
 \alpha_{n+1}\left(\alpha_{n+1}^2-3\beta_{n+1}^2\right).
\ea
Then the coefficients $w_1,\ldots,w_{n+1}$ and the eigenvalues of $W_{-1}^{(1)},\ldots,W_{-1}^{(n)}$ are 
the functions of these conformal weights.
The remaining eigenvalues of $W_{-1}^{(n+1)}$ and $\Wj_{-2}$ ($j=1,\ldots,n+1$) can be written in terms of those of $W_p$ ($p\geq -2$).
By comparing (\ref{def-TW+}) and (\ref{W+Rn}) in the collision limit, or $y\gg z_i$, 
we can read off the eigenvalues of $W_p$ as
\ba\label{Wp}
W_p|R_n\rangle=\left(
 \frac12(p+1)(p+2)z_j^p w_j+(p+2)z_j^{p+1}\Wj_{-1}+z_j^{p+2}\Wj_{-2}
 \right)|R_n\rangle\,,
\ea
then the remaining eigenvalues can be written in terms of those of $W_p$ ($p=-2,\ldots,n-1$):
\ba\label{sub}
\Wi_{-2}|R_n\rangle&=&\sum_{p=0}^{n-1}M_{i,p}\cW_p|R_n\rangle\qquad
(\text{for }i=1,\ldots,n),\nt
W^{(n+1)}_{-2}|R_n\rangle&=&W_{-2}|R_n\rangle+\sum_{p=0}^{n-1}M_{n+1,p}\cW_p|R_n\rangle,\nt
W^{(n+1)}_{-1}|R_n\rangle&=&\cW_{-1}|R_n\rangle+\sum_{p=0}^{n-1}M_{n+2,p}\cW_p|R_n\rangle,
\ea
where 
\ba\label{cWp}
\cW_p&:=&W_p
 -\sum_{i=1}^n (p+2)z_i^{p+1}\Wi_{-1}
 -\sum_{j=1}^{n+1}\frac12 (p+1)(p+2)z_j^pw_j\, , \nt
M_{i,p}&=&\frac{(-1)^{p}}{z_i^2\prod_{j\neq i}(z_j-z_i)}\sum_{\substack{1\leq j_1<\cdots<j_{n-p-1}\leq n\\[+1pt]j_1,\ldots,\,j_{n-p-1}\neq i}} z_{j_1}\cdots z_{j_{n-p-1}}\qquad
(\text{for }i=1,\ldots,n)\, , \nt
M_{n+1,p}&=&\frac{(-1)^{p+1}}{\prod_{i=1}^n z_i^2}
 \sum_{\substack{1\leq j_1<\cdots<j_k\leq n\\[+1pt]
  r_1,\ldots,\,r_k=1,2}}
  z_{j_1}^{r_1}\cdots z_{j_k}^{r_k}
  \quad (\text{where}~r_1+\cdots+r_k=n-p-1)\, ,\nt
M_{n+2,p}&=&\frac{(-1)^{p+1}}{\prod_{i=1}^n z_i}\sum_{1\leq j_1<\cdots<j_{n-p-1}\leq n} z_{j_1}\cdots z_{j_{n-p-1}}\,.
\ea


Now we take the limit (\ref{limit}).
In order to make the discussion clearer, we divide the action of W-current (\ref{W+Rn}) 
into three parts by using the relations in (\ref{sub}):
\ba\label{W}
W_+|R_n\rangle
&=&\left(
\sum_{p=-2}^{n-1}\frac{\zeta_p}{y^{p+3}}W_p
+\sum_{i=1}^n \frac{\xi_i}{y^2}\Wi_{-1}
+\sum_{j=1}^{n+1} \frac{\chi_j}{y^3}w_j
\right)|R_n\rangle
\nt
&=:&
\left(W_+^{(A)}+W_+^{(B)}+W_+^{(C)}\right)|R_n\rangle\, ,
\ea
where $\zeta_p$, $\xi_i$ and $\chi_j$ are the functions of $y$ and $z_i$.


First we read off the coefficients of $W_p$ terms ($p=-2,\ldots,n-1$) as
\ba\label{I}
W_+^{(A)}|R_n\rangle
=\left(\frac{W_{-2}}{y}+\frac{W_{-1}}{y^2}
+\sum_{p=0}^{n-1}\frac{\zeta_p}{y^{p+3}}W_p
\right)|R_n\rangle\, ,
\ea
where
\ba
\zeta_p=
\frac{1}{\prod_{i=1}^{n}(y-z_i)}
\sum_{q=0}^{n-p-1}\sum_{1\leq j_1<\cdots<j_q\leq n}
 (-1)^q z_{j_1}\cdots z_{j_q}\cdot y^{n-q}\,.
\ea
Then in the limit (\ref{limit}), this becomes 
\ba\label{I'}
W_+^{(A)}|R_n\rangle
&\to&
\sum_{p=-2}^{n-1} \frac{W_p}{y^{p+3}}|R_n\rangle\,.
\ea


Next we can similarly read off the coefficients of $\Wi_{-1}$ terms ($i=1,\ldots,n$) as
\ba\label{II}
W_+^{(B)}|R_n\rangle=
\sum_{i=1}^n \frac{z_i^2\prod_{j\neq i}(z_i-z_j)}
{y^2(y-z_i)\prod_{k=1}^n(y-z_k)}\Wi_{-1}|R_n\rangle\, ,
\ea
where $j=1,\ldots,n$. 
By using the degenerate condition (\ref{degenerate}) for $\Wi_{-1}$, we find
\ba
\Wi_{-1}|R_n\rangle = \frac32\sqrt{\kappa}\alp_i\parfrac{}{z_i}|R_n\rangle\,.
\ea
Let us here redefine the state
\ba
|R_n\rangle=:\prod_{1\leq i<j\leq n+1}(z_i-z_j)^{2\alpha_i\alpha_j}|\wtRn\rangle\,,
\ea
just as in (\ref{R2}) for $n=2$.
As we commented there, there is a subtlety  in the choice of overall factor. 
We will argue this issue in section \ref{sec:conclusions},
and here we assume that this normalization properly works.
After this redefinition, 
we divide (\ref{II}) into two parts as
\ba\label{II'}
W_+^{(B)}|\wtRn\rangle
&=&
\frac{3}2\sqrt{\kappa}
\sum_{i=1}^n \frac{z_i^2\prod_{l\neq i}(z_i-z_l)\cdot \alpha_i}{y^2(y-z_i)\prod_{k=1}^n(y-z_k)}\Biggl(
 \parfrac{}{z_i}+\sum_{\substack{1\leq j\leq n+1\\[+1pt]j\neq i}}\frac{2\alpha_i\alpha_j}{z_i-z_j}
\Biggr)|\wtRn\rangle
\nt
&=:&\left(W_+^{(B1)}+W_+^{(B2)}\right)|\wtRn\rangle\,.
\ea

Finally, the coefficients of $w_i$ terms ($i=1,\ldots,n+1$) can be read off as
\ba\label{III}
W_+^{(C)}|\wtRn\rangle
&=& \frac{1}{y^2\prod_{k=1}^n(y-z_k)}
 \sum_{i=1}^{n}\frac{z_i\prod_{l\neq i}(z_i-z_l)}{y-z_i}
\Biggl(
 \sum_{\substack{1\leq j\leq n\\[+1pt]j\neq i}}\frac{z_i}{z_i-z_j}
 +2+\frac{z_i}{y-z_i}\Biggr)
 w_i|\wtRn\rangle
\nt&&
 +\,\frac{\prod_{l=1}^n(-z_l)}{y^3\prod_{k=1}^n(y-z_k)}w_{n+1}|\wtRn\rangle
\ea
where $w_i$ and $w_{n+1}$ are given in (\ref{weight}).

In order to take the limit (\ref{limit}) of these terms, we should note that
\ba\label{limit1}
z_i\prod_{\substack{1\leq j\leq n\\[+1pt]j\neq i}}(z_i-z_j)\cdot\alpha_i
=\sum_{p=0}^n c_p z_i^{n-p}
\,,\quad
z_i\frac{\partial}{\partial z_i}
=z_i\sum_{p=0}^n \frac{\partial c_p}{\partial z_i}\frac{\partial}{\partial c_p}
=\sum_{p=1}^n c_p^{(i)}\parfrac{}{c_p}\,,
\ea
for $i=1,\ldots,n$, where
\ba
c_p^{(i)} :=
(-1)^p \sum_{\substack{1\leq j\leq n+1\\[+1pt]j\neq i}} 
\Biggl(\alpha_j\sum_{\substack{1\leq j_1<\cdots<j_{p-1}\leq n\\[+1pt]j_1,\ldots,\,j_{p-1}\neq i,j}}z_iz_{j_1}\cdots z_{j_{p-1}}\Biggr)\,.
\ea
Therefore, in the limit (\ref{limit}), the terms including the differentials in (\ref{II'}) become
\ba\label{II''}
W_+^{(B1)}|\wtRn\rangle
&=&
\frac32\frac{\sqrt{\kappa}}{y^2\prod_{k=1}^n(y-z_k)}
\sum_{p=1}^n \sum_{q=0}^n \sum_{i=1}^n
\frac{c_p^{(i)}z_i^{n-q}}{y-z_i}c_q\parfrac{}{c_p}
\nt
&\to&
\frac32\frac{\sqrt{\kappa}}{y^{n+3}}
\sum_{p=1}^n \sum_{q=0}^n \sum_{r=0}^{q-p}
 \frac{p\,c_{n+p-q+r}}{y^r}c_q\parfrac{}{c_p}|\wtRn\rangle\,,
\ea
and the remaining terms become
\ba\label{III'}
\left(W_+^{(B2)}+W_+^{(C)}\right)|\wtRn\rangle
\to
 \frac{\sqrt{\kappa}}{y^{n+3}}\Biggl(
 \sum_{\substack{0\leq p,q,r\leq n\\[+1pt]p+q+r\geq n}}
  \frac{c_p c_q c_r}{y^{p+q+r-n}}
 -3c_n(\beta_{n+1}^2+\tfrac18n(n+3)Q^2)\Biggr)|\wtRn\rangle\,.
\nn\\[-10pt]
\ea


To summarize, by putting the results (\ref{I'}), (\ref{II''}) and (\ref{III'}) together, we obtain the final form:
\ba
W_+|\wtRn\rangle
=\left(
\sum_{p=-2}^{n-1}\frac{W_p}{y^{p+3}}
+\sum_{q=n}^{3n}\frac{\sqrt{\kappa}C_q}{y^{q+3}}
\right)|\wtRn\rangle\,,
\ea
where 
\ba
C_q
&=&
\!\sum_{\substack{0\leq r\leq s\leq t\leq n\\[+1pt]r+s+t=q}}
 \frac{3!\, c_r c_s c_t}{(1+\delta_{r,s}+\delta_{s,t})!} 
+\!\sum_{\substack{0\leq r\leq s\leq n\\[+2pt]t=r+s-q\geq 1\\[+1pt](q\leq 2n-1)}}\!
\frac{3t\,c_{r}c_{s}}{1+\delta_{r,s}} \parfrac{}{c_t}
-\delta_{q,n}\cdot 
 3c_n(\beta_{n+1}^2+\tfrac18n(n+3)Q^2)\,.
\nn\\[-17pt]
\ea
Therefore, we can clearly see that the irregular state $|\wtRn\rangle$ is a simultaneous eigenstate of $W_{2n},\ldots,W_{3n}$, the actions of $W_n,\ldots,W_{2n-1}$ on it are given by the first order differential operators, and it is annihilated by the higher modes $W_{k>3n}$. 
However, the actions of $W_0,W_1,\ldots,W_{n-1}$ on the state cannot be found out in our discussion.
{Presumably} this is because in this article 
{we mainly use the information of $\cW_3$ Ward identities and
do not look at the internal momentum dependence carefully.}
We {may miss} 
necessary information to determine the resulting irregular state.
To make the derivation of the irregular state complete, 
we have to deal with the collision limit more precisely by taking the internal channel into account.


\section{Isolated SCFT with $SU(2)$ flavor symmetry}
\label{sec:SU(2)}
  The CFT computations in the previous sections were performed only locally. 
  Namely we considered the collision of the punctures on the open disk around the origin. 
  In order to look at the corresponding $\CN =2$ theories on the gauge theory side, 
  we should add a point at infinity to obtain the Riemann sphere, 
  on which the compactification of the six-dimensional $\CN =(2,0)$ theory is made. 
  The compactification on $C_{0,n+2}$ with a particular marking
  gives a (UV superconformal) linear quiver gauge theory with $n-1$ gauge group factors. 
  Thus the colliding limit of several (regular) punctures on the CFT side 
  corresponds to an appropriate scaling limit of linear quiver gauge theories. 
  
  At a particular locus on the Coulomb branch of $\CN=2$ gauge theory
  where mutually non-local 
  particles become massless, 
  it is known that the theory 
  is an interacting 
  SCFT \cite{Argyres:1995jj}.
  This kind of special points has been found in various papers 
  \cite{Argyres:1995xn, Eguchi:1996vu, Eguchi:1996ds, Gaiotto:2010jf, Seo:2012ns, Giacomelli:2012ea}.
  Also the possible classification was discussed 
  in \cite{Argyres:2005pp, Argyres:2005wx, Argyres:2010py, Chacaltana:2012ch}.
  We then expect that the colliding limit considered on the CFT side is the same 
  as the limit which leads the quiver gauge theory into the nontrivial fixed point.
  In this section, we illustrate this idea by showing how $SU(2)$ linear quiver gauge theory 
  gives the isolated SCFT with an $SU(2)$ flavor symmetry, 
  whose irregular states have been introduced in \cite{Bonelli:2011aa}.
  In the next section, we will apply a similar procedure to $SU(3)$ linear quiver gauge theory
  to obtain isolated SCFT's with an $SU(3)$ flavor symmetry. 
  
  We focus on the colliding limit of $n+1$ regular punctures at the origin while fixing a regular puncture at infinity.
  In the gauge theory view point, this corresponds to the scaling limit to the Argyres-Douglas (AD) fixed points
  of the linear quiver gauge theory:
    \bea
    2 - \underbrace{SU(2) - SU(2) - \cdots - SU(2)
    \vphantom{\begin{pmatrix}\sin \theta \cos \theta \end{pmatrix}}}_{n-1} -\,2
    ,
    \label{D2k}
    \eea
  where each $SU(2)$ represents an $SU(2)$ vector multiplet
  and the number attached to the left or right of the quiver is the number of the fundamentals.
  In section \ref{subsec:D2k}, we will show that the maximal conformal point of this quiver is 
  indeed the $\CW(A_{1}, C_{0,1,\{n+1\}})$ theory.
  Note that this theory can be also obtained as the maximal conformal point of $\SO(4n)$ SYM theory
  (thus it is called as $D_{2n}$ theory)
  or of $SU(2n-1)$ gauge theory with two flavors \cite{Bonelli:2011aa, Argyres:2012fu}.
  The Seiberg-Witten curve of (the relevant deformation of) this SCFT is given by
    \bea
    x^{2}
     =     \frac{1}{w^{2n+2}} + \frac{c_{n}}{w^{2n+1}} + \cdots + \frac{c_{1}}{w^{n+3}}
         + \frac{c_0}{w^{n+2}} + \frac{v_{1}}{w^{n+1}} + \cdots + \frac{v_{n}}{w^{3}} + \frac{m_{-}^{2}}{w^{2}}.
    \eea
  The parameters $v_{i}$ and $c_{i}$ ($i=1, \ldots, n-1$) are, respectively, 
  the VEV's of the relevant deformation operators $V_{i}$ and their corresponding couplings,
  which are added to the Lagrangian by $\delta \CL = \sum_{i} \int d^{2} \theta_{1} d^{2} \theta_{2} c_{i} V_{i}$.
  On the other hand, $c_{0}$ is the mass parameter associated with the $U(1)$ global symmetry.
  The deformation parameters appearing in the Seiberg-Witten curve 
  are always classified in these three types. 
    
  We also note that 
  it is also possible to get the $\CW(A_{1}, C_{0,1,\{ n+1 \}})$ theory starting from a different linear quiver:
    \bea
    1 - \underbrace{SU(2) - SU(2) - \cdots - SU(2)
    \vphantom{\begin{pmatrix}\sin \theta \cos \theta \end{pmatrix}}
    }_{n-1}
     -\,2.
    \label{D2k'}
    \eea
  In other words, the maximal conformal points of the quivers \eqref{D2k} and \eqref{D2k'} are equivalent.
  In the case with $n=2$, this was found in \cite{Argyres:1995xn}.
  
  There is another class of isolated SCFT's $\CW(A_{1}, C_{0,1,\{ n+\frac{1}{2}\}})$.
  One can show that this can be obtained as a sub maximal conformal point of the linear quiver \eqref{D2k}.
  However in section \ref{subsec:D2k-1}, we will look at a different quiver
    \bea
    \underbrace{SU(2) - SU(2) - \cdots - SU(2)
    \vphantom{\begin{pmatrix}\sin \theta \cos \theta \end{pmatrix}}
    }_{n-1}
     -\,2.
    \label{D2k-1}
    \eea
  and show that $\CW(A_{1}, C_{0,1,\{ n+\frac{1}{2}\}})$ is the maximal conformal point of it.
  While this class of SCFT's may not be related with the irregular states constructed by the collision limit 
  in section \ref{sec:Virasoro}, we will study this for completeness.
  Notice however that an explicit expression of the state has been obtained in \cite{Bonelli:2011aa}.

  Let us first write down the Seiberg-Witten curve of the linear quiver \eqref{D2k}.
  We denote the mass parameters of the hypermultiplets 
  on the left of $SU(2)_{1}$ and on the right of $SU(2)_{n-1}$
  as $m_{3,4}$ and $m_{1,2}$ respectively.
  We will define $m_\pm = (m_1\pm m_2)/2$ and $\tilde{m}_{\pm} = (m_{3} \pm m_{4})/2$.
  We also denote the mass parameters of the bifundamentals as $\hat{m}_i$ ($i=1, \ldots, n-2$)
  where the first bifundamental with mass $\hat{m}_1$ is coupled to $SU(2)_1$ and $SU(2)_2$, and so on.
  
  Since each $SU(2)_{i}$ ($i=1, \ldots, n-1$) gauge group is UV superconformal, 
  there are $n-1$ gauge coupling constants $q_{i}$.
  Finally we denote the Coulomb moduli parameters as $u_{i}$ ($i=1, \ldots, n-1$).
  
  The M-theory curve can be written as \cite{Gaiotto:2009we,Witten:1997sc}
    \bea
    (v+m_1)(v+m_2) t^{n+1} + \sum_{i=1}^{n-1} C_i (v^2 + M_i v - u_i) t^{i} + C (v - m_{3}-\mh)(v-m_{4} - \mh)
     = 0, \quad
     \label{kM}
    \eea
  where $C_i$ and $C$ are constants which depend on the coupling constants $q_i$, and
  $\mh:=\sum_{i=1}^{n-2}\mh_i$ is the sum of all the bifundamental mass parameters.
  From the type IIA brane configuration, it is reasonable to have the overall shifts by $\mh$ in the last term.
  $M_{i}$ are unknown constants which will be fixed later.
  We can rewrite the curve as
    \bea
    \prod_{i=1}^{n} (t - t_i) \cdot v^2 + X(t) v + Y(t) 
     =     0,
    \eea
  where we have defined $t_i$ such that 
    \ba\label{t_i}
    \prod_{i=1}^{n}(t-t_i) = t^{n} + \sum_{j=1}^{n-1}C_jt^{j} + C
    \ea
  and 
    \bea
    X(t)
    &=&    2 m_+ t^{n} + \sum_{i=1}^{n-1} C_i M_i t^{i} - 2 C (\tilde{m}_{+} + \hat{m}),
           \nonumber \\
    Y(t)
    &=&    (m_+^2 - m_-^2) t^{n} - \sum_{i=1}^{n-1} C_i u_i t^{i} + C (m_{3} + \hat{m})(m_{4} + \mh).
    \eea
  Note that $C = \prod_{i=1}^{n}(-t_i)$.
  By shifting $v$ to absorb the linear term and defining $v = xt$, we get
    \bea
    x^2
     =     \left( \frac{X(t)}{2t\prod_{i=1}^{n}(t - t_i)} \right)^2
         - \frac{Y(t)}{t^2\prod_{i=1}^{n}(t - t_i)}\,.
    \eea
  The Seiberg-Witten differential in this coordinate is $\lambda_{{\rm SW}} = x dt$.
  It is possible to choose $M_{i}$ in $X$ such that the terms in the parenthesis become
    \bea
    \frac{m_+}{t-t_1} + \sum_{i=2}^{n-1} \frac{t_{i} \hat{m}_{i-1}}{t(t - t_{i})}
         + \frac{t_{n} \tilde{m}_{+}}{t(t-t_{n})} .
    \eea
  Then, after some algebra, we obtain
    \bea
    x^2
    &=&    \left( \frac{m_+}{t-t_1} 
         + \sum_{i=2}^{n-1} \frac{\hat{m}_{i-1}}{t - t_{i}}
         + \frac{\tilde{m}_{+}}{t - t_{n}} + \frac{\tilde{m}_{-}}{t} \right)^2
           \nonumber \\
    & &    ~~~~~~~
         - \frac{\{ (m_{+} + \tilde{m}_+ + \tilde{m}_{-} +\mh)^2 - m_{-}^{2} \} t^{n-1}
         + \sum_{i=1}^{n}C_i \tilde{u}_i t^{i-1}}{t\prod_{i=1}^{n}(t-t_i)}\,,
           \label{kSW}
    \eea
  where $\tilde{u}_{i} = u_{i} + \cdots$.
  It is easy to see that the differential $\lambda_{{\rm SW}} = x dt$ has poles at 
  $t = 0$, $t_{n}$, $t_i$, $t_{1}$ and $\infty$
  whose residues are $\tilde{m}_{-}$, $\tilde{m}_+$, $\hat{m}_{i-1}$, $m_{+}$ and $m_-$.
  This curve is a double cover of the sphere with $n+2$ regular punctures.
  We are free to fix one of $t_i$, so we fix $t_1 = 1$.
  
\subsection{$\CW( A_{1},C_{0,1,\{n+1\} } )$  theory}
\label{subsec:D2k}
  We now consider the maximal degeneration limit of the Seiberg-Witten curve \eqref{kSW},
  which corresponds to the maximal conformal fixed point.
  First of all, let us observe that the curve \eqref{kSW} can be rewritten as
    \bea
    x^{2}
     =     \frac{f_{2n}(t)}{t^{2} \prod_{i=1}^{n}(t - t_{i})^{2}},
           \label{kSW'}
    \eea
  where $f_{2n}(t) = m_{-}^{2} t^{2n} + \cdots + C^{2} \tilde{m}_{-}^{2}$.
  This implies that the branch points of the curve are at the roots of the $2n$-th polynomial $f_{2n}$.
  The genus is $n-1$ agreeing with the number of the Coulomb moduli.
  
  Since we have $2n$ parameters, $\tilde{m}_{-}$, $\tilde{m}_{+}$, $\hat{m}_{i}$, $m_{+}$ and $u_{i}$, 
  the branch cuts can be tuned to be scaled as $(C \tilde{m}_{-})^{1/n}$ as $C \tilde{m}_{-} \rightarrow \infty$.
  As we will soon see below, this corresponds to the maximal degeneration point of the curve.
  In order to focus on this point, we set the coordinate $t$ as $t = (C \tilde{m}_{-})^{a} w$ ($a > 0$).
  By this, we still have a curve of the same genus $n-1$.
  By substituting this into \eqref{kSW'}, we get
    \bea
    x^{2}
     =     \left( \frac{m_{-}^{2}}{w^{2}} + \cdots 
         + \frac{(C\tilde{m}_{-})^{2}}{4 (C\tilde{m}_{-})^{2na} w^{2n+2}} \right)
           \prod_{i=1}^{n-1}\left(1 - \frac{t_{i}}{(C\tilde{m}_{-})^{a} w}\right)^{-2}.
           \label{SWcurvelinear}
    \eea
  Note that we have multiplied the r.h.s.~by $(C\tilde{m}_{-})^{2a}$ 
  since we are considering the quadratic differential $x^{2} (dt)^{2}$.
  Note also that the $1/w^{2n+2}$ term is the highest one such that the curve is of genus $n-1$.
  This determines $a = 1/n$.
  
\subsubsection*{$n=2$ case}
\label{subsubsec:k=2}
  In order to illustrate how we can take this limit more precisely, let us first consider $n=2$ case.
  The original gauge theory is simply $SU(2)$ with four flavors 
  where $m_1$, $m_2$, $m_{3}$ and $m_{4}$ are the mass parameters of hypers, $u$ is the Coulomb moduli.
  In this case the curve is
    \bea
    x^2
     =     \left( \frac{m_+}{t-1} + \frac{\tilde{m}_{+}}{t-q} + \frac{\tilde{m}_{-}}{t} \right)^2
         - \frac{\{(\tilde{m}_{-}+\tilde{m}_+ + m_{+})^2 - m_-^2\} t + \tilde{u}}{t(t-1)(t-q)}.
           \label{k=2SW}
    \eea
  The Seiberg-Witten differential has three poles at
  $t = 0$, $q$, $1$ and $\infty$ with residues $\tilde{m}_{-}$, $\tilde{m}_+$, $m_{+}$ and $m_-$ respectively.
  Note that $C=q$ in this case.
  
  It is useful to rewrite the terms in the r.h.s.~of the parenthesis in \eqref{k=2SW} as
    \bea
    \frac{g_{2}(t)}{t(t-1)(t-q)},
    \eea
  where $g_{2}$ is 
    \bea
    g_{2}(t)
     =     (\tilde{m}_{-} + \tilde{m}_{+} + m_{+}) t^{2} - ((1+q)\tilde{m}_{-} + \tilde{m}_{+} + qm_{+}) t
         + q \tilde{m}_{-}.
    \eea
  
  We now consider the limit where the punctures at $t=q$ and $t=1$ collide to the one at $t=0$
  while the puncture at $t=\infty$ is fixed.
  This means that we fix the mass parameter $m_-$.
  Then we want to find the limit by scaling $\tilde{m}_{-}$, $\tilde{m}_+$, $m_{+}$ and $\tilde{u}$
  such that the Seiberg-Witten curve maximally degenerates.
  At the same time, we have to scale the local coordinate of the sphere $t$ as
  $t = (q\tilde{m}_{-})^{1/2} w$, as we noticed above.
  Let us first consider the first term in the r.h.s.~of \eqref{k=2SW}.
  This can be written as
    \bea
    \left( \frac{g_{2}(t)}{(q\tilde{m}_{-}) w^{3}} \right)^{2} 
    \left(1-\frac{1}{(q\tilde{m}_{-})^{1/2}w} \right)^{-2} \left(1-\frac{q}{(q\tilde{m}_{-})^{1/2}w} \right)^{-2},
     \label{parenthesis}
    \eea
  Again notice that we have multiplied the r.h.s.~by the overall factor $qm_{-}$.
  Thus, we demand that these are finite in the limit $C\tilde{m}_{-} \to \infty$.
  This fixes that 
    \bea
    \tilde{m}_{-}+\tilde{m}_+ + m_{+}
     =:     c_0, ~~~~
    - \frac{(1+q) \tilde{m}_{-} + \tilde{m}_{+} + q m_{+}}{(q\tilde{m}_{-})^{1/2}}
     =:     c_1,
           \label{k=2limit}
    \eea
  which completely determine the scaling of $m_{+}$ and $\tilde{m}_{+}$.
  Thus we get
    \bea
    \left( \frac{c_0}{w} + \frac{c_1}{w^{2}} + \frac{1}{w^{3}} \right)^{2}.
    \label{parenthesis2}
    \eea
  Notice that it is impossible to have higher order terms in $w$ with keeping all the terms finite.
  (If possible, the genus of the curve could be greater than that of the original quiver.)  
    
  Then, we consider the last term in the r.h.s.~of \eqref{k=2SW}.
  The first term stays finite combining with $\frac{(\tilde{m}_{-}+\tilde{m}_+ + m_{+})^{2}}{w^{2}}$ 
  coming from \eqref{parenthesis2}, and the second term is expanded as 
  $- \frac{\tilde{u}}{(q\tilde{m}_{-})^{1/2} w^{3}} + \cdots$.
  It is impossible to have the higher order finite terms and therefore the scaling of $\tilde{u}$ is
    \bea
    \tilde{u}
     =   - (q\tilde{m}_{-})^{1/2} v.
           \label{k=2Coulomb}
    \eea
  These fix the scaling of all the parameters and finally we get
    \bea
    x^{2}
    &=&    \left( \frac{1}{w^{3}} + \frac{c_1}{w^{2}} + \frac{c_0}{w} \right)^{2} 
         - \frac{c_0^2 - m_-^2}{w^2} + \frac{v}{w^{3}}
           \nonumber \\
    &=&    \frac{1}{w^{6}} + \frac{2c_1}{w^{5}} + \frac{2c_0+c_1^{2}}{w^{4}}
         + \frac{v + 2c_1 c_0}{w^{3}} + \frac{m_{-}^{2}}{w^{2}},
    \eea
  which is the Seiberg-Witten curve of the $\CW(A_{1}, C_{0,1,\{ 3 \}})$ theory.
    
  Note that we treated the first and second terms in \eqref{k=2SW} separately, when we considered the limit.
  Indeed, this is the only way to produce the most singular pole at $t=0$.
  On the CFT side, we may focus only on the limit of the external momenta and the complex structures
  which is the same as the one \eqref{k=2limit} where the first term in \eqref{k=2SW} is finite.
  The CFT side is implicit for the variable corresponding to the Coulomb moduli.

\subsubsection*{Generic $n$} 

  We now consider the limit where $n$ regular punctures collide to the one at $t = 0$.
  We fix the puncture at $t = \infty$ with residue $m_-$ unchanged.
  As in the $n=2$ case, let us focus on the terms in the parenthesis in the r.h.s.~of \eqref{kSW}
  which can be written as
    \bea
    \frac{g_{n}(t)}{t \prod_{i=1}^{n}(t-t_{i})},
    \eea
  where $g_{n}$ is an $n$-th polynomial:
    \bea
    g_{n}
     =     c_0 t^{n} + \sum_{i=1}^{n-1} \hat{c}_{i} t^{n-i} + C \tilde{m}_{-}.
    \eea
  We have defined 
    \bea
    c_0
    &=&    \tilde{m}_{+} + \tilde{m} + \mh,
           \\
    \hat{c}_{1}
    &=&  - t_{1} (\tilde{m}_{-} + \tilde{m}_{+} + \mh)
         - \sum_{i=2}^{n-1} t_{i} (\tilde{m}_{+} + \tilde{m}_{-} + m_{+} + \mh - \mh_{i-1})
         - t_{n} (\tilde{m}_{-} + m_{+} + \hat{m}),
           \nonumber 
    \eea
  and so on.
  Note that $f_{2n} = g_{n}^{2} + \cdots$.
  By scaling the coordinate $t = (C\tilde{m}_{-})^{1/n} w$, the first term in the r.h.s.~of \eqref{kSW} is
    \bea
    & &    \Bigg( \frac{c_0}{w} + \sum_{i=1}^{n-1} \frac{\hat{c}_{i}}{(C\tilde{m}_{-})^{i/n} w^{i+1}}
         + \frac{1}{w^{n+1}} \Bigg)^2.
           \label{kSWscale}
    \eea
  Thus, we keep 
    \bea
    c_0 ~~~{\rm and}~~~
    \frac{\hat{c}_i}{(C\tilde{m}_{-})^{i/n}}
     =: c_{i}
    & &    \label{scalek}
    \eea
  finite, 
  where $i=1, \ldots, n-1$.
  Finally, by appropriately scaling $\tilde{u}_i$ ($i=1, \ldots, n-1$), we get from the last term in \eqref{kSW}
    \bea
    \frac{v_{n}}{w^3} + \cdots + \frac{v_{1}}{w^{n+1}}.
    \eea
  By combining these altogether, we obtain 
    \bea
    x^2
     =     \left( \frac{1}{w^{n+1}} + \sum_{i=1}^{n-1} \frac{c_i}{w^{i+1}} + \frac{c_0}{w} \right)^2
         - \frac{c_0 - m_-^2}{w^2} + \sum_{j=1}^{n-1} \frac{v_{n-j}}{w^{j+2}},
    \eea
  which is the Seiberg-Witten curve of the $\CW(A_{1}, C_{0,1,\{ n+1\}})$ theory, 
  a double cover of a sphere with one irregular puncture at $t=0$ of degree $n+1$ 
  and one regular puncture at $t = \infty$.
  The residues of them are $c_0$ and $m_-$ respectively.

\subsection{$\CW (A_{1}, C_{0,1,\{ n+\frac{1}{2}\}})$ theory}
\label{subsec:D2k-1}
  As we commented, it is possible to obtain the $\CW(A_{1}, C_{0,1,\{ n+\frac{1}{2}\}})$ theory 
  as a sub maximal conformal point of the same quiver \eqref{D2k}.
  Instead of doing so, we will proceed two steps here:
  we obtain the curve of the reduced quiver \eqref{D2k-1} and then consider the maximal conformal point of it.
  
  The first step can be done similarly to the calculation in the previous section. 
  The Seiberg-Witten curve of the quiver \eqref{D2k-1} is
    \bea
    x^{2}
    &=&    \left( \frac{m_{+}}{t-t_{1}} + \sum_{i=2}^{n-1} \frac{\hat{m}_{i-1}}{t - t_{i}} \right)^{2}
         - \frac{(m_+ + \mh)^2 - m_-^2}{t^2}
         + \frac{\Lambda^{2}}{t^{3}}
         + \frac{\sum_{i=1}^{n-1}C_i \tilde{u}_i t^{i-1}}{t^2 \prod_{i=1}^{n-1}(t-t_i)}\,.
    \eea
  We will set $t_{1}$ as $t_{1} = 1$ below.
  Note that it is possible to obtain this curve from \eqref{SWcurvelinear} 
  by taking the limit decoupling the massive flavors:
  the collision of two punctures at $t=0$ and $t_{n}$ giving the irregular puncture at $t=0$.
  
  The curve can be rewritten as
    \bea
    x^{2}
     =     \frac{f_{2n-1}(t)}{t^{3} \prod_{i=1}^{n-1}(t - t_{i})^{2}},
    \eea
  where 
    \bea
    f_{2n-1}
     =     m_{-}^{2} t^{2n-1} + \cdots + (C\Lambda)^{2}.
    \eea
  The branch points are at the roots of the $(2n-1)$-th polynomial $f_{2n-1}$ and at $t=0$.
  Thus the genus is $n-1$.

  In order to obtain the maximal conformal point of this quiver, we take the limit $(C\Lambda) \to \infty$, 
  as in the previous subsection.
  In this limit, the branch cuts are at $t=0$ and at $t = \CO((C\Lambda)^{\frac{1}{2n-1}})$.
  In order to focus on the physics around the latter region, we scale the coordinate 
  as $t = (C\Lambda)^{a} w$ with $a>0$.
  Note that the resulting Seiberg-Witten curve should be of genus $n$.
  It follows that the curve is written as
    \bea
    x^{2}
    &=&    \Bigg( \frac{m_{-}^{2}}{w^{2}} 
         + \sum_{i=1}^{n-1} \frac{\hat{v}_{n-i}}{(C\Lambda)^{ia} w^{i+2}}
         + \sum_{i=1}^{n-1} \frac{\hat{c}_{i}}{(C \Lambda)^{(n+i-1)a} w^{n+i+1}}
           \nonumber \\
    & &     ~~~~~~~~~~~~~~~~~~~
         + \frac{(C\Lambda)^{2}}{(C\Lambda)^{(2n-1)a} w^{2n+1}} \Bigg)
           \prod_{i=1}^{n-1}\left(1 - \frac{t_{i}}{(C\Lambda)^{a} w}\right)^{-2},
    \eea
  where $\hat{v}_{i}$ and $\hat{c}_{i}$ are combinations of the mass parameters and the Coulomb moduli.
  It follows from the genus of the curve is $n-1$ that $a = \frac{2}{2n-1}$. 
  Then, it is possible to keep the combinations
    \bea
    \frac{\hat{v}_{n-i}}{(C\Lambda)^{ia}}
     =:
           v_{n-i}, ~~~~
    \frac{\hat{c}_{i}}{(C \Lambda)^{(n+i-1)a}}
     =:
           c_{i}
    \eea
  finite.
  Thus, the resulting curve is 
    \bea
    x^{2}
     =     \frac{m_{-}^{2}}{w^{2}} + \sum_{i=1}^{n-1} \frac{v_{n-i}}{w^{i+2}}
         + \sum_{i=1}^{n-1} \frac{c_{i}}{w^{n+i+1}} + \frac{1}{w^{2n+1}}.
    \eea

\subsubsection*{$n=2$ case}
  For illustration of the limit we consider the case of $n=2$. 
  Namely, the $SU(2)$ gauge theory with two flavors scaling to the $\CW(A_{1}, C_{0,1,\{ \frac{5}{2} \}})$ theory as the maximal conformal point.
  The curve in this case is
    \bea
    x^{2}
     =     \frac{m_{+}^{2}}{(t-1)^{2}} - \frac{m_+^2 - m_-^2}{t^2}
         + \frac{\Lambda^{2}}{t^{3}}
         - \frac{\tilde{u}}{t^2 (t - 1)}
    \eea
  We can rewrite this as
    \bea
    x^{2}
     =     \frac{f_{3}(t)}{t^{3}(t-1)^{2}}
    \eea
  where
    \bea
    f_{3}(t)
     =     m_{-}^{2} t^{3} + (2m_{+}^{2} - 2m_{-}^{2} - \tilde{u} + \Lambda^{2}) t^{2} 
         + (- m_{+}^{2} + m_{-}^{2} + \tilde{u}-2\Lambda^{2})t + \Lambda^{2}.
    \eea
  
  In order to get the maximal conformal point, we scale $t= (-\Lambda)^{2/3}$ where $C=-1$
  with the following combinations being fixed
    \bea
    \frac{2m_{+}^{2} - 2m_{-}^{2} - \tilde{u} + \Lambda^{2}}{(- \Lambda)^{2/3}}
     =:
           v, ~~~~
    \frac{-m_{+}^{2} + m_{-}^{2} + \tilde{u}-2\Lambda^{2}}{(-\Lambda)^{4/3}}
     =:
           c.
    \eea
  Note that this prescription of the limit is slightly different from the on 
  in the $N = 2n$ case where we consider the limit of the terms separately.
  The solution is
    \bea
    m_{+}^{2}
     =     \Lambda^{2} + (-\Lambda)^{4/3} c + (-\Lambda)^{2/3} v, ~~~
    \tilde{u}
     =     3 \Lambda^{2} + 2 (-\Lambda)^{4/3} c + (-\Lambda)^{2/3} v.
    \eea
    Therefore, we get the curve
    \bea
    x^{2}
     =     \frac{m_{-}^{2}}{w^{2}} + \frac{v}{w^{3}} + \frac{c}{w^{4}} + \frac{1}{w^{5}}
    \eea
  which is one of the $\CW (A_{1}, C_{0,1,\{ \frac{5}{2}\}})$ theories.


\section{Isolated SCFT with $SU(3)$ flavor symmetry}
\label{sec:SU(3)}
Now we consider the isolated SCFT with an $SU(3)$ flavor symmetry
by generalizing the argument in the previous section.
Recently, such kind of SCFT's has been found 
by string theory consideration \cite{Cecotti:2010fi}, the BPS quiver method \cite{Cecotti:2012jx}
(see also \cite{Alim:2011ae, Alim:2011kw})
and the correspondence with the Hitchin system {and 3d mirror symmetry \cite{Xie:2012hs, XZ}} 
(see also \cite{Nanopoulos:2010zb}).
These are generalizations of the one first found in \cite{Eguchi:1996vu, Eguchi:1996ds}
as an IR fixed point of $SU(3)$ SQCD.

The $\CW_{3}$ irregular states constructed in section \ref{sec:W3} indicate
that there should be a series of SCFT's, $\CW(A_{2}, C_{0,1,\{ n\}})$ theory, associated with $C_{0,1,\{ n\}}$, 
where the degree of the irregular puncture is counted with respect to the Seiberg-Witten differential. 
In section \ref{subsec:1}, we find they arise from $SU(3)$ linear quiver gauge theories 
as a nontrivial IR fixed point on the Coulomb branch.
In section \ref{subsec:SU(3)other}, we study {isolated} SCFT's from other quiver gauge theories,
though the relation with the two-dimensional CFT is not clear.
In section \ref{subsec:comparison}, 
we {compare the $\CW(A_{2}, C_{0,1,\{ n \}})$ theories with} the ones studied in \cite{Cecotti:2012jx}.

\subsection{$\CW(A_{2}, C_{0,1,\{ n+1\}})$ theory}
\label{subsec:1}
Let us first consider the $SU(3)^{n-1}$ gauge theory with $3+3$ fundamental hypermultiplets
\ba
3-
\underbrace{SU(3) - SU(3) - \cdots - SU(3)
 \vphantom{\begin{pmatrix}\sin \theta \cos \theta \end{pmatrix}}}_{n-1}
-\,3.
\label{quiver:3+3}
\ea
The following computation goes in parallel with
section \ref{subsec:D2k}.
The M-theory curve is
\ba\label{curve}
\prod_{a=1,2,3} (v+m_a)\cdot t^{n}
+\sum_{i=1}^{n-1} C_i(v^3 + P_i v^2 + Q_i v + R_i)t^{i}
+C\prod_{b=4,5,6}(v-(m_b+\mh))=0, ~~~~~
\ea
where $\mh$ is the sum of bifundamental mass parameters $\mh_j$ ($j=1,\ldots,n-2$),
and $P_{i}$, $Q_{i}$ and $R_{i}$ are functions of them and the Coulomb moduli.
$C_i$ and $C$ are determined by the positions of punctures $t=t_i$ ($i=1,\ldots,n$) 
in the same way as (\ref{t_i}).
Then we can rewrite (\ref{curve}) as
\ba
\prod_{i=1}^{n}(t-t_i)\cdot v^3+X(t)v^2+Y(t)v+Z(t)=0,
\ea
where
\ba\label{XYZ}
X(t) &=& 3m_+ t^{n} + \sum_{i=1}^{n-1} C_iP_i t^{i} - 3C(\tm_+ + \mh), \nt
Y(t) &=& M_2 t^{n} + \sum_{i=1}^{n-1} C_iQ_i t^{i} + C\tilde M_2, \nt
Z(t) &=& m_1m_2m_3 t^{n} + \sum_{i=1}^{n-1} C_iR_i t^{i} - C(m_4+\mh)(m_5+\mh)(m_6+\mh)\,.
\ea
Here we define $m_+ = \frac13(m_1+m_2+m_3)$, $\tm_+ = \frac13(m_4+m_5+m_6)$,
and 
\ba
M_2 = \sum_{1\leq i<j\leq 3} m_i m_j\,,\quad
\tilde M_2 = \sum_{4\leq i<j\leq 6} (m_i+\mh)(m_j+\mh)\,.
\ea
By shifting $v$ to eliminate the $v^2$ terms and defining $x:={v}/{t}$, we obtain 
\ba
x^3+\phi^{(2)}(t)\,x+\phi^{(3)}(t)=0.
\ea
The quadratic and the cubic differentials, $\phi^{(2)}$ and $\phi^{(3)}$, are
\ba\label{phi2-1}
\phi^{(2)}(t)&=&
-3 \left(\frac{X}{3t\prod_{i=1}^{n}(t-t_i)}\right)^2
+\frac{Y}{t^2\prod_{i=1}^{n}(t-t_i)}\,,
\\\label{phi3-1}
\phi^{(3)}(t)&=&
2\left(\frac{X}{3t\prod_{i=1}^{n}(t-t_i)}\right)^3
-\frac{XY}{3t^3\prod_{i=1}^{n}(t-t_i)^2}
+\frac{Z}{t^3\prod_{i=1}^{n}(t-t_i)}\,.
\ea
In the following, we set $t_1=1$ and $1>t_2>\cdots>t_{n}>0$.


We demand that the Seiberg-Witten differential $\lambda_{\textrm{SW}}=xdt$
has a regular pole of simple type at $t=1$ with residue $(2m_+,-m_+,-m_+)$. 
Similarly, there are poles of the same type at $t=t_2,\ldots,t_n$, 
and their residues must be of the same form with $m_+$ replaced 
by $\mh_1,\ldots,\mh_{n-2}$ or $\tm_+$, respectively.
This means that $X(t)$ satisfies
\ba
\frac{X(t)}{3t\prod_{i=1}^{n}(t-t_i)}
=\frac{m_+}{t-1}+\sum_{j=2}^{n-1}\frac{t_j\mh_{j-1}}{t(t-t_j)}+\frac{t_{n}\tm_+}{t(t-t_{n})}\,,
\ea
{which determines $P_{i}$ as}
\ba
P_i&=&\frac{3}{C_i}(-1)^{n-i}
\!\sum_{1\leq p_1<\cdots<p_{n-i}\leq n} t_{p_1}\cdots t_{p_{n-i}}\!
\nt&&\qquad\qquad\qquad\qquad\qquad\quad\times
\left(
m_+(1-\delta_{p_1,1}) - \textstyle{\sum_{a=1}^{n-i}}\mh_{p_a-1} - \tm_+\delta_{p_{n-i},n}
\right).
\ea
%

{Then, after some algebra, we can write the quadratic differential (\ref{phi2-1}) as}
\ba\label{phi2}
\phi^{(2)}(t)
=-3\left(
\frac{m_+}{t-1}
+\sum_{i=2}^{n-1}\frac{\mh_{i-1}}{t-t_i}
+\frac{\tm_+}{t-t_{n}}+\frac{\tilde m_-}{t}
\right)^2
+\frac{V_2{\:\!}t^{n-1}+\sum_{j=1}^{n-1} C_ju_j^{(2)}t^{j-1}}{t\prod_{k=1}^{n}(t-t_k)},
\quad
\ea
where $\eps:=\tm_++\tm_-+\mh$,
\ba
\label{V2}
V_2 = M_2 + 6m_+ \eps + 3\eps^2\,,
\quad
u_j^{(2)} = Q_j+2P_j\eps+3\eps^2\,,
\ea
and also we defined
\ba
-3\tm_-^2\ =: -3 \tilde m_+^2 + \sum_{4\leq i<j\leq 6}m_im_j = \tilde M_2 - 3(\tilde m_++\mh)^2,
\ea
Note that the $\tilde{m}_{-}^{2}$ depends only on $m_4$, $m_5$ and $m_6$.
Similarly, the cubic differential (\ref{phi3-1}) can be rewritten as
\ba
\label{phi3}
\phi^{(3)}(t)
&=&2\left(
\frac{m_+}{t-1}
+\sum_{i=2}^{n-1}\frac{\mh_{i-1}}{t-t_i}
+\frac{\tm_+}{t-t_{n}}+\frac{\tm_-}{t}
\right)^3
\nt&&
+\frac{V_4}{t^2\prod_{k=1}^{n}(t-t_k)^2}
+\frac{V_3{\:\!}t^{n}+\sum_{j=1}^{n-1} C_j u_j^{(3)}t^{j}+C\hat M_3}{t^3\prod_{k=1}^{n}(t-t_k)}\,,
\quad
\ea
where 
\ba
\label{V3}
V_{3}&=&m_1m_2m_3 -3m_+\eps^2 -2\eps^3
\,,\quad
u_j^{(3)}=
R_j -P_j \eps^2 -2\eps^3
\,,\nt
V_4&=& -\frac13X(t)
\left(V_2{\:\!}t^{n-1}+\sum_{j=1}^{n-1} C_ju_j^{(2)}t^{j-1}\right)
=-m_+V_2{\:\!}t^{2n-1}+\cdots\,,\nt
\Mh_3&=&-(m_4+\mh)(m_5+\mh)(m_6+\mh)+3(\tm_++\mh)\eps^2-2\eps^3\,.
\ea

Let us check the residues of the Seiberg-Witten differential at the poles $t=\infty$ and $0$.
In the limit of $t\to\infty$, the leading terms in $\phi^{(2)}$ and $\phi^{(3)}$ are $\frac{M_2-3m_+^2}{t^2}$ 
and $(m_{1}m_{2}m_{3} - m_{+} M_{2} + 2 m_{+}^{3})/t^3$, respectively.
Therefore we find that at $t=\infty$ there is a regular puncture of full type whose residue is 
\ba
\frac13(-2m_1+m_2+m_3,~ m_1-2m_2+m_3,~ m_1+m_2-2m_3)\,.
\ea
Similarly in the limit of $t\to 0$, the leading terms in $\phi^{(2)}$ and $\phi^{(3)}$ are
{$-\frac{3\tilde{m}_{-}^{2}}{t^2}$} 
and {$(-m_4m_{5}m_{6} + \tilde{m}_{+}\sum_{4 \leq i < j \leq 6} m_{i} m_{j} - 2 \tilde{m}_{+}^{3})/t^{3}$}, respectively.
Thus we find that the puncture at $t=0$ is also regular full type one whose residue is
\ba
\frac13(2m_4-m_5-m_6,~ -m_4+2m_5-m_6, ~-m_4-m_5+2m_6)\,.
\ea


Now we consider the limit where the $n$ punctures at 
$t=1,t_2,\ldots,t_{n}$ simultaneously 
collide to the puncture at $t=0$, while the remaining puncture at $t=\infty$ is kept intact.
In order to take such a limit, let us rescale the coordinates as
\ba\label{rescale}
t=(C\tm_-)^{1/n} w
\ea
and take the limit $C\tm_-\to \infty$ with suitable variables kept finite.
We first consider 
\ba
\frac{m_+}{t-1}+\sum_{i=2}^{n-1} \frac{\mh_{i-1}}{t-t_i}+\frac{\tm_+}{t-t_{n}}+\frac{\tm_-}{t}
=:
\frac{c_0t^{n}+\sum_{j=1}^{n-1}\hc_jt^{n-j}+C\tm_-}{t\prod_{k=1}^{n}(t-t_k)}
\ea 
in the differentials (\ref{phi2}) and (\ref{phi3}).
Here we define
\ba\label{cond1}
c_0&:=&m_+ + \tm_+ +\tm_- + \mh \nt
\hc_j&:=&(-1)^j
\sum_{1\leq p_1<\cdots<p_j\leq n} t_{p_1}\cdots t_{p_j}
\nt&&\,\times
\Bigl(m_+(1-\delta_{p_1,1})+(\mh-\textstyle{\sum_{a=1}^j \mh_{p_a-1}})+\tm_+(1-\delta_{p_j,n})+\tm_-\Bigr)\,.
\ea
Then we require that the following variables should be kept finite in the limit:
\ba\label{5.21}
c_0\,,\quad
c_j:=\frac{\hc_j}{(C\tm_-)^{j/n}}\,,
\ea
where $j=1,\ldots,n-1$.
Note that all of the variables available here are $n$ mass parameters, so we can have at most $n$ finite parameters.

In order to find the final form of the quadratic and the cubic differentials,
we must also keep the following variables finite:
\ba\label{5.22}
\frac{C_ju_j^{(2)}}{(C\tm_-)^{1-j/n}}
&=&jv_{j}^{(2)}+\sum_{\ell=1}^{j-1}\ell c_{n-j+\ell} v_{\ell}^{(2)}
,
\nt
\frac{C_j\hat u_j^{(3)}}{(C\tm_-)^{1-j/n}}
&=&v_{n-j}^{(3)}
\,,\quad
\frac{\Mh_3+V_2(\tm_++\mh)}{\tm_-}=\beta^2,
\quad
\ea
where we define
$\hat u_j^{(3)} := u_j^{(3)} + \eps u_j^{(2)} -\frac13 V_2P_j$.

Therefore we finally find in the limit the quadratic differential becomes
\ba\label{phi2-f}
\phi^{(2)}=
-3\left(\frac{1}{w^{n+1}}+\sum_{i=1}^{n-1}\frac{c_i}{w^{i+1}}+\frac{c_0}{w}\right)^2
+\sum_{j=1}^{n-1}\frac{(n-j)v_{n-j}^{(2)}+\sum_{p=1}^{n-j-1} p\,c_{j+p}v_p^{(2)}}{w^{j+2}}
+\frac{V_2}{w^2}
\quad~
\ea
and the cubic differential becomes
\ba
\phi^{(3)}&=&
2\left(\frac{1}{w^{n+1}}+\sum_{i=1}^{n-1}\frac{c_i}{w^{i+1}}+\frac{c_0}{w}\right)^3
\nt&&
+\sum_{j=1}^{2n-1}\!\sum_{\substack{1\leq p\leq n\\[+1pt]1\leq q\leq n-j\\[+1pt]0\leq j-p+q\leq n}}\! \frac{q\,c_p c_{j-p+q} v_q^{(2)}}{w^{j+3}}
+\frac{\beta^2}{w^{n+3}}
+\sum_{\ell=1}^{n-1}\frac{v_\ell^{(3)}}{w^{\ell+3}}
+\frac{V_3-m_+V_2}{w^3}~,
\nn\\[-17pt]
\ea
where
$V_2$ and $V_3$ are defined in (\ref{V2}) and (\ref{V3}),
and we set $c_n=1$.
It is easy to see that the Seiberg-Witten differential has a simple pole at $t=\infty$ of full type
and a pole of degree $n+1$ at $t=0$. 
Thus, this theory is associated with $C_{0,1,\{ n+1 \} }$.
  
  The scaling dimensions of the parameters can be calculated by demanding that the Seiberg-Witten differential
  has dimension one.
  Since $x^{3} + \phi_{2} x + \phi_{3} = 0$, this completely fixes the dimensions of the parameters in the differentials (\ref{5.21}) and (\ref{5.22}) as
    \bea
    \Delta(c_{i})
    &=&    1 - \frac{i}{n}, ~~~
    \Delta(c_{0})
     =
    \Delta(\beta)
     =     1, ~~~
    \Delta(V_{2})
     =     2,~~~
    \Delta(V_{3})
     =     3.
           \nonumber \\
    \Delta(v_{i}^{(2)})
    &=&    1 + \frac{i}{n},~~~
    \Delta(v_{i}^{(3)})
     =     3 - \frac{i}{n},
    \label{SU(3)quiver}
    \eea
  where $i=1, \ldots, n-1$.
  We therefore see that the parameters $c_{i}$ and $v_{i}^{(2)}$ are paired to give the relevant deformations.
  On the other hand, $v_{i}^{(3)}$ have dimensions greater than two, 
  thus they are interpreted as irrelevant operators.
  {Notice that this SCFT is not the most general one with the leading singularity of order $n+1$ at $w=0$, 
  {\it i.e.} $\lambda_{{\rm SW}} \sim \frac{dw}{w^{n+1}}$,
  since $\phi_{2}$ and $\phi_{3}$ include common parameters $c_{i}$ and $v_{i}^{(2)}$. 
  This corresponds to the degenerate case studied 
  in \cite{Xie:2012hs, XZ}\footnote{We would like to thank Dan Xie for helpful explanation of his works.}.
  Generically it is possible {for} 
  $\phi_{2}$ and $\phi_{3}$ {to have} 
  independent parameters.
  This is {a} different SCFT from the one we obtained here, 
  and might be given as an IR fixed point of $\CN=2$ $SU(3)$ gauge theory with non-Lagrangian $E_{6}$ SCFT's
  associated with a sphere where all the punctures are of full type.
  }

By comparing the results in section \ref{sec:W3} and this subsection,
and assuming the correspondence between the cubic differential of gauge theory and the W-current of Toda theory 
\ba
\phi^{(3)} \to \frac{2}{\sqrt{\kappa}}W_+\,,
\ea
we find the correspondence of the parameters in both theories is 
\ba
c_0 \to c_0\,,\quad c_i\to c_i\,,\quad c_n \to 1\,,\quad
v_j^{(2)} \to 3\frac{\partial}{\partial c_j}\,,\quad
\beta^2 \to -6\left(\beta_{n+1}^2+\tfrac18 n(n+3)Q^2\right).
\quad
\ea
We cannot see the counterparts of $v_j^{(3)}$ in the Toda theory, 
since the actions of $W_0,\ldots,W_{n-1}$ on the irregular state are not found out in our discussion.

Finally we note that the same isolated SCFT $\cW(A_2,C_{0,1,\{n+1\}})$ can be also obtained 
from a different quiver gauge theory:
\ba
2-
\underbrace{SU(3) - SU(3) - \cdots - SU(3)
 \vphantom{\begin{pmatrix}\sin \theta \cos \theta \end{pmatrix}}}_{n-1}
- 3.
\ea
It means that the maximal conformal point of this quiver is the same as that of (\ref{quiver:3+3}).

\subsection{Other quiver theories}
\label{subsec:SU(3)other}
As in section \ref{subsec:D2k-1}, we can consider other classes of quiver gauge theories:
\ba
1 - \underbrace{SU(3) - SU(3) - \cdots - SU(3)
 \vphantom{\begin{pmatrix}\sin \theta \cos \theta \end{pmatrix}}}_{n-1}
-\,3,
\ea
and
\ba\label{SU(3)asymquiver}
\underbrace{SU(3) - SU(3) - \cdots - SU(3)
 \vphantom{\begin{pmatrix}\sin \theta \cos \theta \end{pmatrix}}}_{n-1}
-\,3,
\ea
to search isolated SCFT's with an $SU(3)$ flavor symmetry.
Since the relation of these SCFT's with the irregular state in $\CW_{3}$ algebra is unclear,
we will not {pursue general cases}.
Instead, we consider only $n=2,3$ for the quiver \eqref{SU(3)asymquiver}.

Let us derive the Seiberg-Witten curve of the quiver theory \eqref{SU(3)asymquiver}.
The M-theory curve is
\ba
\prod_{a=1,2,3}(v+m_a)\cdot t^n
+\sum_{i=1}^{n-1} C_i(v^3+P_iv^2+Q_iv+R_i)t^i+(C\Lambda)^3=0\,,
\ea
where $\Lambda$ is the dynamical scale.\footnote{
We take $m_{4,5,6}\to \infty$ and $t_n\to 0$ with $t_nm_4m_5m_6=\Lambda^3$ kept finite in the previous case (\ref{curve}).}
We set $\prod_{k=1}^{n-1} (t-t_k) = t^{n-1}+\sum_{i=1}^{n-1} C_it^{i-1}$ and $C=C_1^{1/3}$.
By a similar calculation to the previous case, we obtain the quadratic and the cubic differentials as 
\ba
\phi^{(2)}(t)
&=&-3\left(
\frac{m_+}{t-1}
+\sum_{i=2}^{n-1}\frac{\mh_{i-1}}{t-t_i}
\right)^2
+\frac{V_2{\:\!}t^{n-1}+\sum_{j=1}^{n-1} C_ju_j^{(2)}t^{j-1}}{t^2\prod_{k=1}^{n-1}(t-t_k)}\,,
\\
\phi^{(3)}(t)
&=&2\left(
\frac{m_+}{t-1}
+\sum_{i=2}^{n-1}\frac{\mh_{i-1}}{t-t_i}
\right)^3
+\frac{V_4}{t^3\prod_{k=1}^{n-1}(t-t_k)^2}
+\frac{V_3{\:\!}t^{n}+\sum_{j=1}^{n-1} C_j u_j^{(3)}t^{j}+(C\Lambda)^3}{t^4\prod_{k=1}^{n-1}(t-t_k)}\,,
\nn
\ea
and
\ba\label{V-u}
V_2 &=& M_2 + 6m_+ \mh + 3\mh^2\,,
\quad
u_j^{(2)} = Q_j+2P_j\mh+3\mh^2\,.\nt
V_{3}&=&m_1m_2m_3 -3m_+\mh^2 -2\mh^3
\,,\quad
u_j^{(3)}=
R_j -P_j \mh^2 -2\mh^3
\,,\nt
V_4&=& -\frac{X(t)}{3t}
\left(V_2{\:\!}t^{n-1}+\sum_{j=1}^{n-1} C_ju_j^{(2)}t^{j-1}\right)
=-m_+V_2{\:\!}t^{2n-2}+\cdots\,.
\ea
The Seiberg-Witten differential has regular poles of simple type at $t=t_{i}$ ($i=1, \ldots, n-1$)
and of full type at $t=\infty$.
The pole at $t=0$ is of irregular whose degree is $\frac{4}{3}$.

The maximal conformal point of this quiver can be obtained by taking the limit where the $n-1$ regular punctures
collide to the irregular one at $t=0$.
In order to take such a limit, we rescale the coordinate as
\ba
t=(C\Lambda)^{a}w
\ea
and then take the limit $C\Lambda\to \infty$ with some variables kept finite. 

\subsubsection*{$n=2$ case}

  We now consider the maximal conformal point of the $n=2$ case, 
  namely $SU(3)$ SQCD with $N_{f}=3$.
  We first consider the quadratic differential $\phi^{(2)}$ whose expansion is 
\bea
  \phi^{(2)}(w)
  = \frac{V_2 - 3m_{+}^{2}}{w^{2}}
    + \frac{C_1u_1^{(2)}-6m_{+}^{2}+V_2}{(C\Lambda)^{a} w^{3}}  
    + \frac{C_1u_1^{(2)}-9m_{+}^{2}+V_2}{(C\Lambda)^{2a} w^{4}} + \cdots\,.
  \label{phi2phi2}
\eea
  Let $m_{+}$ be a finite parameter here. 
  It follows that the second term can be kept finite by $C_1u_1^{(2)} = (C\Lambda)^{a} v_1^{(2)}$,
  where $v_1^{(2)}$ is a finite parameter, and the higher order terms are suppressed.
  After taking the limit, we get
    \bea
    \phi^{(2)}(w)
     = \frac{v_1^{(2)}}{w^{3}} + \frac{V_{2} - 3m_{+}^{2}}{w^{2}}\,.
    \eea
  We next consider the cubic differential $\phi^{(3)}$ whose expansion is
\bea
  \phi^{(3)}(w)
  &=& \frac{V_3 - m_+V_2 + 2m_{+}^{3}}{w^{3}}
  + \frac{6m_{+}^{3} + C_1u_1^{(3)} - m_+C_1u_1^{(2)}-2m_+V_2+V_3}{(C\Lambda)^{a} w^{4}}
  \nt&&  
  + \frac{12m_{+}^{3} + C_1u_1^{(3)} - 2m_+C_1u_1^{(2)}-3m_+V_2+V_3 +(C\Lambda)^3}{(C\Lambda)^{2a} w^{5}} + \cdots.
  \label{phi3phi3}
\eea
  Since $m_{+}$ is finite and $C_1u_1^{(2)} \sim (C\Lambda)^{a}$, in order to make the $1/w^4$ and $1/w^{5}$ terms finite, we need to set
  $C_1u_1^{(3)} = (v_1^{(3)}+m_+v_1^{(2)})(C\Lambda)^{a}$,
  where $v_1^{(3)}$ is a {second} finite parameter, and $a = 3/2$.
  After taking the limit, we get
    \bea
    \phi^{(3)}(w)
     = \frac{1}{w^{5}} + \frac{v_1^{(3)}}{w^{4}} 
     + \frac{V_{3} - m_+V_{2} + 2m_{+}^{3}}{w^{3}}\,.
    \eea 
Thus we obtain the Seiberg-Witten curve of the $\cW(A_2,C_{0,1,\{ \frac{5}{3} \}})$ theory.

The dimensions of the parameters are 
\ba
\Delta(m_+)=1\,,\quad
\Delta(V_2)=2\,,\quad
\Delta(V_3)=3\,,\quad
\Delta(v_1^{(2)})=\frac12\,,\quad
\Delta(v_1^{(3)})=\frac32\,,
\ea
which agree with the ones of the class 2 SCFT of $SU(3)$ with $N_{f} = 3$ in \cite{Eguchi:1996vu}. 
  Note that the $\CW(A_{1}, C_{0,1,\{ 2 \}})$ theory (or the $D_{4}$ theory) studied in section \ref{subsec:D2k} 
  has similar deformation parameters, except for $m_{+}$.
  (As discussed in \cite{Eguchi:1996vu}, this mass parameter $m_{+}$ can be freely tuned.) 
  Thus we conclude that $\CW(A_{2}, C_{0,1,\{ \frac{5}{3} \}}) = \CW(A_{1}, C_{0,1,\{ 2 \}})$.

\subsubsection*{$n=3$ case}

For the calculation of maximal conformal point in the $n=3$ case, 
it is convenient to define the variables $c_0$, $\hc_1$ as
\ba\label{c0c1}
\frac{m_+}{t-1}+\frac{\mh_1}{t-t_2}
=:\frac{c_0t+\hat c_1}{(t-1)(t-t_2)}\,.
\ea
The quadratic differential is expanded as
\ba
\phi^{(2)}(w)&=&\frac{V_2-3c_0^2}{w^2}
+\frac{-6c_0\hc_1+C_2u_2^{(2)}}{(C\Lambda)^aw^3}
+\frac{-3\hc_1^2+C_1u_1^{(2)}+C_2u_2^{(2)}}{(C\Lambda)^{2a}w^4}
\nt&&
+\frac{-6\hc_1^2+C_1u_1^{(2)}+C_2u_2^{(2)}}{(C\Lambda)^{3a}w^5}
+\cdots\,.
\ea
Here we assume that $V_2$ and $c_0$ are finite, and drop the vanishing terms in the collision limit.
Let us impose the following conditions:
\ba
&&\back
\hc_1=c_1(C\Lambda)^a+\zeta(C\Lambda)^{4a/3}\,,
\nt&&\back
C_2u_2^{(2)}=(v^{(2)}+6c_0c_1)(C\Lambda)^a+6c_0\zeta(C\Lambda)^{4a/3}\,,
\nt&&\back
C_1u_1^{(2)}=(c^{(2)}+3c_1^2)(C\Lambda)^{2a}+6c_1\zeta(C\Lambda)^{7a/3}+3\zeta^2(C\Lambda)^{8a/3}\,,
\ea
where $c_1$, $\zeta$, $c^{(2)}$ and $v^{(2)}$ are kept finite.
After taking the limit we obtain 
\ba
\phi^{(2)}(w)=
\frac{c^{(2)}}{w^4}+\frac{v^{(2)}}{w^3}+\frac{V_2-3c_0^2}{w^2}\,.
\ea
On the other hand, the cubic differential is expanded as 
\ba
\phi^{(3)}(w)
&=&\frac{V_3-m_+V_2+2c_0^3}{w^3}
+\frac{6c_0^2\hc_1-m_+C_2u_2^{(2)}+C_2u_2^{(3)}}{(C\Lambda)^aw^4}
\nt&&
+\frac{6c_0\hc_1^2-m_+C_1u_1^{(2)}-2m_+C_2u_2^{(2)}+C_1u_1^{(3)}+C_2u_2^{(3)}}{(C\Lambda)^{2a}w^5}
\nt&&
+\frac{2\hc_1^3-2m_+C_1u_1^{(2)}-3m_+C_2u_2^{(2)}+C_1u_1^{(3)}+C_2u_2^{(3)}+(C\Lambda)^3}{(C\Lambda)^{3a}w^6}
\nt&&
+\frac{6\hc_1^3-3m_+C_1u_1^{(2)}-4m_+C_2u_2^{(2)}+C_1u_1^{(3)}+C_2u_2^{(3)}+(C\Lambda)^3}{(C\Lambda)^{4a}w^7}
+\cdots\,,
\ea
where we assume that $V_3-m_+V_2$ remains finite.
We note that $m_+=(c_0+\hc_1)/(1-t_2)=\zeta(C\Lambda)^{4a/3}+\cdots$ from (\ref{c0c1}).
In order to make the $1/w^4$ and $1/w^5$ terms finite,
we need to impose the following conditions:
\ba
C_2u_2^{(3)}&=&(v_2^{(3)}-6c_0^2c_1)(C\Lambda)^a+\cdots+(v^{(2)}+6c_0c_1)\zeta(C\Lambda)^{7a/3}+6c_0\zeta^2(C\Lambda)^{8a/3}\,,
\nt
C_1u_1^{(3)}&=&(v_1^{(3)}-6c_0c_1^2)(C\Lambda)^{2a}+\cdots
\nt&&\qquad\quad~
\cdots+(c^{(2)}+3c_1^2)\zeta(C\Lambda)^{10a/3}+6c_1\zeta^2(C\Lambda)^{11a/3}+3\zeta^3(C\Lambda)^{4a}\,.
\ea
To make the $1/w^6$ term finite, we need to impose additional conditions
\ba
a=\frac34\,,\quad
\zeta=1\,,\quad
c^{(2)}=3c_1^2\,.
\ea
Then the $1/w^7$ term also becomes finite. 
After taking the limit we obtain
\ba
\phi^{(3)}=\frac{1}{w^7}+\frac{c^{(3)}}{w^6}+\frac{v_1^{(3)}}{w^5}+\frac{v_2^{(3)}}{w^4}+\frac{V_3-m_+V_2+2c_0^3}{w^3}\,,
\ea
where we define $c^{(3)}=2c_1^3$.
Thus we obtain the Seiberg-Witten curve of the $\cW(A_2,C_{0,1,\{\frac73\}})$ theory.

The dimensions of the parameters are 
\ba
&&\back
\Delta(c_0)=1\,,\quad
\Delta(V_2)=2\,,\quad
\Delta(V_3-m_+V_2)=3\,,
\nt&&\back
\Delta(c^{(2)})=\frac12\,,\quad
\Delta(v^{(2)})=\frac54\,,\quad
\Delta(c^{(3)})=\frac34\,,\quad
\Delta(v_1^{(3)})=\frac32\,,\quad
\Delta(v_2^{(3)})=\frac94\,.
\ea
Therefore, the pairs of parameters $(c^{(2)}, v_1^{(3)})$ and $(c^{(3)}, v^{(2)})$ give the relevant deformations, 
and the parameter $v_2^{(3)}$ is interpreted as the irrelevant operator.
{This isolated SCFT, however, seems unusual because} 
both the relevant couplings $c^{(2)}$ and $c^{(3)}$ can be written in terms of $c_{1}$.
This means that 
{they} are simultaneously turned on 
and cannot be independently shifted in this case.


\subsection{Comparison with BPS quiver method}
\label{subsec:comparison}
  In \cite{Cecotti:2012jx}, isolated SCFT's denoted by $D(G, n)$
  with a flavor symmetry $G$ have been considered,
  based on type IIB compactification 
  on local Calabi-Yau three-fold specified by $\hat{A}(s,t) \times A_{p}$.
  (The defining equation of this will be given shortly.)
  The four-dimensional theory can be considered as $\CN=2$ $SU(p+1)$ gauge theory
  coupled to two strongly coupled sectors labeled by $s$ and $t$.
  By decoupling the gauge sector of this theory 
  one ends up with decoupled SCFT's $D(SU(p+1), s-1)$ and $D(SU(p+1), t-1)$ each with a global symmetry $SU(p+1)$
  which should be (a subgroup of) a flavor symmetry of the SCFT.
  
  {In this subsection, we compare the {SCFT's} found in previous subsections with the $D(SU(3), n)$ SCFT.}
  To do this, let us consider the Calabi-Yau geometry of $\hat{A}(s,t) \times A_{p}$
  which is described by the equation
    \bea
    W
     =  \Lambda^b(   z^{s} + z^{-t} )+ P_{p+1}(x) + y^{2} + w^{2} =0,
    \eea
  where $b$ is the coefficient of the one-loop beta function of the gauge coupling
   and $P_{p+1} = x^{p+1} + u_{2} x^{p-1} + \cdots$.
  Note that by setting $z = e^{z'}$ we recover the expression in \cite{Cecotti:2012jx}.
  The decoupling of the gauge group leads to the geometry of $D(SU(p+1), s-1)$: 
    \bea
    \tilde{W}
     =     z^{s} + P_{p+1}(x) + y^{2} + w^{2} =0,
    \eea 
  where the moduli, $u_{2}, \ldots$, in $P$ become the mass parameters.
  
  The holomorphic three-form of the Calabi-Yau three-fold can be written as
    \bea
    \Omega
     =     \frac{dz}{z} \wedge \left( \frac{dx \wedge dy}{\partial \tilde{W}/\partial w} \right)
     =     \frac{dz}{z} \wedge \left( \frac{dx \wedge dy}{2w} \right).
    \eea
  In order to find the Seiberg-Witten curve and the differential we need to integrate $\Omega$ over the two-spheres,
  parametrized by $const + x^{p+1} + y^{2} + w^{2} = 0$;
  see the calculation in \cite{Klemm:1996bj}.
  After integration we obtain
    \bea
    \lambda_{{\rm SW}}
     =     x \frac{dz}{z},
    \eea
  where $x$ is determined by the curve 
    \bea
    W_{{\rm SW}}
     =     z^{s} + P_{p+1}(x) =0.
    \eea

  First of all, let us check that this is indeed the correct one when $p=1$ namely $D(SU(2), s-1)$ theory.
  In this case, we obtain the curve
    \bea
    W_{{\rm SW}}
     =     z^{s} + x^{2} + u_{2}
     =     0.
    \eea
  By shifting $x \rightarrow x z$ to absorb the $1/z$ factor in the differential, we get
    \bea
    z^{s-2} + x^{2} + \frac{u_{2}}{z^{2}}
     =     0.
    \eea
  By $z \rightarrow 1/t$ we finally obtain
    \bea
    x^{2} + \frac{1}{t^{s+2}} + \frac{u_{2}}{t^{2}}
     =     0. \label{thecurve}
    \eea
  Supplying the less singular terms corresponding to the relevant and the mass deformations,
  {we can make \eqref{thecurve}} 
 identical to the curve of the $\CW(A_{1}, C_{0,1,\{ \frac{s}{2}+1\}})$ theory.

  Then, let us consider the $D(SU(3), s-1)$ theory.
  Repeating the same argument, we obtain the curve
    \bea
    x^{3} + \frac{u_{2}}{t^{2}} x + \frac{1}{t^{s+3}} + \frac{u_{3}}{t^{3}}
     =     0,
    \eea
  with the differential
    \bea
    \lambda_{{\rm SW}}
     =     x dt.
    \eea
  Note that at $t=\infty$ the differential has a simple pole of full type, since $u_{2}$ and $u_{3}$
  are independent mass parameters, and that {the degree of the pole at $t=0$} is $\frac{s}{3} + 1$. 
  {The pole structure of the curve agrees with that of the $\CW(A_{2}, C_{0,1,\{ \frac{s}{3} +1 \}})$ theory.
  The $s=3n$ case corresponds to the SCFT which we found in subsection \ref{subsec:1}.
  However, the pole structure is not enough to identify these SCFT's:
  relevant deformations should be compared. 
  We cannot conclude this point here, 
  since we are not sure that whether the $D(SU(3), s-1)$ theory is non-degenerate or degenerate 
  in the sense that we mentioned at the end of subsection \ref{subsec:1}.
  }
  Furthermore, the cases of $s=2$ and $4$ correspond to $n=2$ and $3$ of the SCFT's 
  found in subsection \ref{subsec:SU(3)other}, respectively.
  {However, in the same {reason} as above, the $n=3$ SCFT might be different from the $D(SU(3), 3)$ theory.
  }


\section{Discussions}
\label{sec:conclusions}

As we have seen in this paper the irregular state for
isolated SCFT's with an $SU(3)$ flavor symmetry is 
a simultaneous eigenvector of the higher positive modes
$L_n, \ldots, L_{2n}$ and $W_{2n}, \ldots, W_{3n}$ with $n \geq 2$.
Some of the lower positive modes act as the first order differential 
operators. We should mention that these conditions 
cannot determine the irregular state in the Verma module uniquely.
In fact this issue already appears in the Virasoro case. 
Let us illustrate it by the simplest example of the Virasoro
irregular state with $n=2$, where we have
\beqa
L_k \vert I_2 \rangle &=& \lambda_k  \vert I_2 \rangle, \qquad (2 \leq k \leq 4), \label{eigenvalue} \\
L_\ell \vert I_2 \rangle &=& 0, \qquad ( 5 \leq \ell). \label{vanishing}
\eeqa
The action of $L_0$ and $L_1$ may be given by some first order differential
operators, which we will discuss later. 
The condition \eqref{vanishing} implies that only non-vanishing
inner products with the basis of the Verma module are
\beq
\langle \Delta, c \vert L_1^{k_1} L_2^{k_2}
L_3^{k_3} L_4^{k_4} \vert I_2 \rangle = 
\lambda_2^{k_2} \lambda_3^{k_3} \lambda_4^{k_4}
\langle \Delta, c \vert L_1^{k_1} \vert I_2 \rangle,
\eeq
where $\vert  \Delta, c \rangle$ is the primary state with the conformal dimension 
$\Delta$ and the central charge $c$.
Hence the irregular state $\vert I_2 \rangle$ is expanded as follows:
\beq
\vert I_2 \rangle = \sum_{k=0}^\infty a_k \vert \Psi_k \rangle,
\eeq
where $ \vert \Psi_k \rangle$ is a vector in the Verma module at level $k$ and
\beq
a_k := \langle \Delta, c \vert L_1^{k} \vert I_2 \rangle.
\eeq
As in the $n=1$ case, or the cases of $\CW(A_1, C_{0,1,\{\frac{3}{2} \}})$ and $\CW(A_1, C_{0,1,\{ 2 \}})$,
the family of states $\vert \Psi_k \rangle$ is completely fixed, once we specify the simultaneous eigenvalues
$\lambda_2, \lambda_3$ and $\lambda_4$. However, there remain an infinite number of 
arbitrary constants $a_n$.
We can expect that the $L_1$ action as a differential operator provides
some recursion relation on $a_n$. In terms of the \lq\lq CFT" parameters $c_{0,1,2}$ associated 
with the collision of three primaries, the eigenvalues are given by
$\lambda_4 =  c_2^2$, $\lambda_3 = 2 c_1 c_2$ and $\lambda_2 =  2 c_0 c_2 + c_1^2$.
Then the action of $L_1$ is
\beq
L_1 \vert I_2 (c_i) \rangle =~\left( v_1 + 2 c_0 c_1 - c_2 \frac{\partial}{\partial c_1} \right)\vert I_2 (c_i) \rangle, 
\eeq
where in addition to $c_{0,1,2}$ we have introduced the parameter $v_1$ coming from the Coulomb moduli on the gauge theory side. 
Using
\beq
v + 2 c_0 c_1 - c_2 \frac{\partial}{\partial c_1} =  e^{\chi(c_i)}  \left( - c_2 \frac{\partial}{\partial c_1} \right) e^{-\chi(c_i)}
\eeq
with $\chi(c_i) :=  \frac{c_0 c_1^2}{2 c_2} + \frac{v c_1}{c_2} $,
we obtain a recursion relation
\beqa
a_n &=& \langle \Delta, c \vert  e^{\chi(c_i)} 
 \left( - c_2 \frac{\partial}{\partial c_1} \right)^n   e^{-\chi(c_i)} \vert I_2 (c_i) \rangle \CR
 &=& e^{\chi(c_i)} 
 \left( - c_2 \frac{\partial}{\partial c_1} \right)^n e^{-\chi(c_i)} a_0 ( c_i ),
\eeqa
where we have used the fact that the primary state $\vert \Delta, c \rangle$ is independent of $c_i$. 
The factor in front of $a_0$ is essentially the Hermite polynomial. 
Hence the total ambiguity in the solutions to the Virasoro irregular state with $n=2$
is the \lq\lq initial" condition $a_0(c_i)$,
once we fix the Verma module (or the \lq\lq final" conformal weight $\Delta (\alpha) $) the irregular state 
$\vert I_2 (c_i) \rangle$ belong to. The last condition is related to the $L_0$ action on $\vert I_2 (c_i) \rangle$.

The above ambiguity of the \lq\lq initial" condition is nothing but 
the overall coefficient ambiguity $a_0$ of the ansatz for the solution
\begin{align}
\label{ansatzI}
\vert I_n\rangle=
a_0(c_i)\big(~\vert \Delta\rangle+\textrm{descendants}~\big).
\end{align}
Since the irregular state conditions involve the derivatives with respect to $c_i$,
the coefficient $a_0$ of the leading term $\vert \Delta\rangle$ may affect the higher-level terms drastically
when one solve the condition recursively. {Fortunately in the simplest case of above, we can derive
the recursion relation. However, in general it is not clear at all that the recursion relations 
coming from the differential operator are under our control.}
This problem is also related to the ansatz to define the normalized state $\vert\widetilde{R_n}\rangle$
from  $\vert {R_n}\rangle$, {which we have used to eliminate infinities in taking the collision limit of punctures. }
In this paper we  {are sloppy}
with the internal momentum dependence of the state $\vert {R_n}\rangle$.
The  {more precise} regular state is
\begin{align}
\vert {R_n};\alpha_i,\beta_j\rangle
=
{\bf{1}}_{\Delta}~ V_{\Delta_1}(z_1)~{\bf{1}}_{\Delta(\beta_1)}
~ V_{\Delta_2}(z_2)~{\bf{1}}_{\Delta(\beta_2)}\cdots
{\bf{1}}_{\Delta(\beta_{n-1})}~ V_{\Delta_n}(z_n)~\vert \Delta_{n+1}\rangle,
\end{align}
where $\beta_i$ is an internal momentum of this channel
and  {${\bf{1}}_{\Delta}$ is the projection operator to 
the Verma module with the corresponding conformal dimension.}
This expression also fixes the ordering of the fusions of the vertex operators
to describe the linear quiver on the dual gauge theory side. 
We see the coefficient of the leading term of $\vert\widetilde{R_n}\rangle$ is
\begin{align}
\vert\widetilde{R_n}\rangle
=\frac{
\langle{\Delta}\vert~ V_{\Delta_1}(z_1)~{\bf{1}}_{\Delta(\beta_1)}
~ V_{\Delta_2}(z_2)~{\bf{1}}_{\Delta(\beta_2)}\cdots
{\bf{1}}_{\Delta(\beta_{n-1})}~ V_{\Delta_n}(z_n)~\vert \Delta_{n+1}\rangle
}{\prod_{i<j}(z_i-z_j)^{2\alpha_i\alpha_j}}
\vert \Delta\rangle+\textrm{descendants}~
,
\end{align}
where this leading coefficient is the fraction between
the $n+2$ point Liouville conformal block
and the free field conformal block.
The behavior of the fraction in the collision limit is not obvious
and one may be afraid of its vanishing as $z_i\to0$.
However, we can define the overall normalization $a_0$
by multiplying it by a certain function $f(\alpha_i)$ of $\alpha_i$
to obtain the non-zero limit
\begin{align}
a_0(c_i,\beta_j)
=\lim_{\textrm{collision}}f(\alpha_i)\frac{
\langle{\Delta}\vert~ V_{\Delta_1}(z_1)~{\bf{1}}_{\Delta(\beta_1)}
~ V_{\Delta_2}(z_2)~{\bf{1}}_{\Delta(\beta_2)}\cdots
{\bf{1}}_{\Delta(\beta_{n-1})}~ V_{\Delta_n}(z_n)~\vert \Delta_{n+1}\rangle
}{\prod_{i<j}(z_i-z_j)^{2\alpha_i\alpha_j}}
.
\end{align}
This coefficient may provide  {a correct normalization (\ref{ansatzI})
for $\vert I_n\rangle$ 
to reproduce the Nekrasov partition function} of the corresponding four-dimensional theory
as a function of $c_i$ and $\beta_j$.
However, since the fraction of the conformal blocks looks a complicated form,
it is very hard to evaluate the limit-value $a_0$ explicitly.

{
In summary, it is an important problem to fix the overall normalization of the irregular state
for working out the AGT-like correspondence of isolated ${\mathcal N}=2$  SCFT. 
The fact that}
this overall coefficient plays a key role to establish the AGT relation
for Pestun's partition functions on $S^4$ \cite{Pestun:2007rz}
{may provide a clue for the problem.}
Let us demonstrate it by the simplest case $n=1$. 
In this case there is essentially no ambiguity of solution
because we can replace the derivative term $\Lambda\partial_\Lambda$ by $L_0$.
Here we assume $Q=0$ for simplicity.
The fraction of conformal blocks in this case 
is merely $z_1^{\Delta-c_0^2}$, and so the overall coefficient is
defined by
\begin{align}
C
=\lim_{\textrm{collision}}\alpha_1^{\Delta-c_0^2}z_1^{\Delta-c_0^2}
=c_1^{\Delta-c_0^2}.
\end{align}
This coefficient provides the classical part $\Lambda^{a^2}$
and a part of proportionality coefficient of the AGT relation
\begin{align}
\textrm{correlation function on }S^4
\propto \int a^2da
(\textrm{DOZZ part})
\vert \Lambda^{a^2}Z^{\textrm{inst}}\vert^2.
\end{align}
This idea may be used to establish a possible AGT relation for isolated  SCFT's with irregular states.

\section*{Acknowledgments}
The authors would like to thank P.~C.~Argyres, H.~Awata, S.~Giacomelli, Y.~Nakayama, 
Y.~Tachikawa, D.~Xie and Y.~Yamada for comments and discussions. 
K.M.~would like to thank the hospitality of KEK theory group.
S.S.~would like to thank the hospitality of the theoretical particle physics group at SISSA.
 The work of H.K. is supported in part by Grant-in-Aid for Scientific Research
(Nos. 22224001 and 24540265) and JSPS Bilateral Joint Projects
 (JSPS-RFBR collaboration) from MEXT, Japan. 
 K.M.~is supported by JSPS postdoctoral fellowships for research abroad.
S.S.~is partially supported by Grant-in-Aid for JSPS fellows (No.\,23-7749).


\appendix

\section{Convention for $A_2$ Toda field theory}
\label{sec:Toda}

The action of two-dimensional $A_2$ Toda field theory is 
\ba
S=\int d^2\sigma\,\sqrt{g}\left(
\frac{1}{8\pi}g^{xy}\partial_x\vvphi\cdot \partial_y\vvphi
+\mu\sum_{k=1,2}e^{b\ve_k\cdot\vvphi}+\frac{Q}{4\pi}R\vrho\cdot\vvphi\right)
\ea
where $\vvphi$ is the Toda fields satisfying $\vvphi\cdot(1,1,1)=0$.
$g_{xy}$ is the metric on 2-dim Riemann surface, and $R$ is its curvature.
$\mu$ is the scale parameter, $b$ is the dimensionless coupling constant, and $Q:=b+1/b$.
$\ve_k$ is the $k$-th simple root and  
$\vrho$ is the Weyl vector ({\em i.e.}~half the sum of all positive roots) of $\mathfrak{sl}_3$ algebra. 

Our convention of $\mathfrak{sl}_3$ algebra is as follows;
Let $\vu_i$ be the orthonormal bases of $\mathbb{R}^3$ with $\vu_i\cdot \vu_j = \delta_{ij}$.
The simple roots are defined as 
\ba
\ve_1 = \vu_1 - \vu_2\,,\quad \ve_2 = \vu_2 - \vu_3\,.
\ea
Together with the maximal root $\vtheta = \ve_1 + \ve_2 = \vu_1 - \vu_3$, 
they form a positive root system of $\mathfrak{sl}_3$. 
The fundamental weights $\vw_1$ and $\vw_2$ are defined by $\vw_i \cdot \ve_j = \delta_{ij}$ and given by
\ba
\vw_1 = \frac{1}{3} (2 \vu_1 - \vu_2 -\vu_3)\,,\quad
\vw_2 = \frac{1}{3} ( \vu_1 + \vu_2 - 2 \vu_3)\,.
\ea
The Weyl vector is $\vrho = \frac{1}{2}(\ve_1 + \ve_2 + \vtheta) = \vw_1 + \vw_2 = \vtheta$,
the last equality is specific to $\mathfrak{sl}_3$ algebra. Finally the weights of the fundamental representation are
\ba
&&\back
\vlam_1=\vw_1 = \frac{1}{3}(2\vu_1 - \vu_2 - \vu_3)\,, \quad
\vlam_2=\vw_1 - \ve_1 = \frac{1}{3} ( -\vu_1 + 2\vu_2 - \vu_3)\,,
\nt&&\back
\vlam_3=\vw_1 - \ve_1 - \ve_2 = \frac{1}{3} (-\vu_1 -\vu_2 + 2 \vu_3)\,.
\ea
In the following, we simply choose $\vu_i$ as $\vu_1=(1,0,0)$, $\vu_2=(0,1,0)$ and $\vu_3=(0,0,1)$.

The symmetry algebra of $A_2$ Toda theory is well known as $W_3$ algebra. The generators of this algebra are defined by the two chiral Noether currents with spin 2 and 3 as
\ba
T(z)=\sum_{n=-\infty}^\infty \frac{L_n}{z^{n+2}}\,,\quad
W(z)=\sum_{n=-\infty}^\infty \frac{W_n}{z^{n+3}}\,.
\ea
The commutation relation among these generators is
\ba
{}[L_n,L_m]&=&(n-m)L_{n+m}+\frac{c}{12}n(n^2-1)\delta_{n+m,0}\nt
{}[L_n,W_m]&=&(2n-m)W_{n+m}\nt
{}\frac29[W_n,W_m]&=&\frac{c}{3\cdot 5!}n(n^2-1)(n^2-4)\delta_{n+m,0}
+\frac{16}{22+5c}(n-m)\Lambda_{n+m}
\nt&&
+(n-m)\left(\frac{1}{15}(n+m+2)(n+m+3)-\frac16(n+2)(m+2)\right)L_{n+m}\quad
\ea
where the central charge is $c=2-24Q^2$ and 
\ba
\Lambda_n=\sum_{k=-\infty}^\infty\!:\!L_kL_{n-k}\!:+\,\frac15 x_nL_n\,;\quad
x_{2l}=(1+l)(1-l)\,,\quad
x_{2l+1}=(2+l)(1-l)\,.
\ea
Note that here we fix the normalization of the generators.
In this convention, all the generators are hermite, {\em i.e.}~the adjoint of generators are $L_n^\dagger=L_{-n}$ and $W_n^\dagger=W_{-n}$.

The highest weight state in this algebra is given by the vertex operator in Toda theory:
\ba
V_\valp(z)=\;:\!e^{\valp\cdot\vvphi(z)}\!:\;,\quad
|V_\valp\rangle=\lim_{z\to 0}V_\valp(z)|0\rangle\,,
\ea
where $\valp\in \bbC^3$ and $\valp\cdot(1,1,1)=0$.
Note that $\valp$ is called Toda momentum, whose concrete form can be determined by the degenerate state condition~\cite{Kanno:2009ga}. Its expression for all the cases in AGT relation is given in \cite{Shiba:2011ya}.\footnote{
The expression in \cite{Shiba:2011ya} has been justified only in the correspondence to the 1-loop partition function of the corresponding gauge theories. The correspondence to the instanton partition function remains a challenging discussion.
For $A_2$ Toda theory, it has been checked in \cite{Kanno:2010kj} up to instanton level 3.
For a general $A_N$ case, the check is still incomplete: The discussion using Heisenberg algebra seems promising~\cite{Fateev:2011hq}, and some researchers suggest $\cW_{1+\infty}$ algebra is useful for this discussion~\cite{Kanno:2011qv}.}
In the maintext of this paper, Toda momentum is denoted as $(\alp_1,\alp_2)$, which means
\ba
\valp&=&-i\left(\frac{\alp_1}{\sqrt{3}}+\frac{Q}{2}\right)(1,1,-2)
 -i\left(\alp_2+\frac{Q}{2}\right)(1,-1,0)
\nt
&=&-i\frac{\alp_1}{\sqrt{3}}(1,1,-2)-i\alp_2(1,-1,0)-iQ\vrho\,.
\ea

The conformal weights of the vertex operator are given as 
\ba
L_0|V_\valp\rangle=\Delta_\valp|V_\valp\rangle\,,\quad
W_0|V_\valp\rangle=w_\valp|V_\valp\rangle\,,\quad
\ea
where 
\ba
\Delta_\valp&=&\frac12(-2iQ\vrho-\valp)\cdot\valp
\,=\,\alp_1^2+\alp_2^2-Q^2 ,\nt
w_\valp&=&i\frac{3}{\sqrt{2}}\sqrt{\frac{48}{22+5c}}\prod_{i=1}^3(\valp+iQ\vrho)\cdot\vlam_i
=
\frac{2}{\sqrt{4-15Q^2}}\,\alp_1(\alp_1^2-3\alp_2^2)\,.
\ea

Finally we show the free field representation of the chiral currents:
\ba
T(z)&=&-\frac12:\!(\partial_z\vvphi)^2\!:-\,iQ\vrho\cdot\partial_z^2\vvphi\,,\\
\frac{\sqrt{2}}{3}W(z)&=&\;:\!\prod_{i=1}^3(\vlam_i\cdot\partial_z\vvphi)\!:
+\,\frac{iQ}{2}\!:\!\left[(\vlam_1\cdot\partial\vvphi)(\ve_1\cdot\partial^2\vvphi)+(\vlam_3\cdot\partial\vvphi)(\ve_2\cdot\partial^2\vvphi)\right]\!:+\,\frac12Q^2\vlam_2\cdot\partial^3\vvphi\,.
\nn
\ea
Similarly to Toda momentum, Toda field can be also denoted as $(\varphi_1,\varphi_2)$, which means
\ba
\vvphi=
 \frac{i}{\sqrt{6}}\varphi_1(1,1,-2)+\frac{i}{\sqrt{2}}\varphi_2(1,-1,0)\,.
\ea
In this notation, the generators are
\ba
T(z)&=&\frac12\left[(\partial_z\varphi_1)^2+(\partial_z\varphi_2)^2\right]
 +\frac{Q}{\sqrt{2}}(\sqrt{3}\partial_z^2\varphi_1+\partial_z^2\varphi_2)
\nt
W(z)&=&\frac{i}{2\sqrt{3}}\left[(\partial_z\varphi_1)^3-3\partial_z\varphi_1(\partial_z\varphi_2)^2\right]
+\frac{\sqrt{3}iQ}{2\sqrt{2}}\left[\partial_z\varphi_1(\sqrt{3}\partial_z^2\varphi_1-2\partial_z^2\varphi_2)-\sqrt{3}\partial_z\varphi_2\partial_z^2\varphi_2\right]
\nt&&
+\,\frac{\sqrt{3}iQ^2}4(\partial_z^3\varphi_1-\sqrt{3}\partial_z^3\varphi_2)
\ea
and the vertex is
$V_\valp(z)=\;:\!e^{\valp\cdot\vvphi}\!:\;
=\;:\!e^{\sqrt{2}(\alp_1\varphi_1+\alp_2\varphi_2)+\frac{Q}{\sqrt{2}}(\sqrt{3}\varphi_1+\varphi_2)}\!:\,$.

\section{Virasoro irregular conformal blocks}
\label{sec:BMTGT}
  Virasoro irregular states which describe degree $\frac{3}{2}$ and $2$ singularities \cite{Gaiotto:2009ma} 
  were shown to be extended to any order in \cite{Bonelli:2011aa, Gaiotto:2012sf, Felinska:2011tn}.
  As classified in \cite{Felinska:2011tn}, the conditions satisfied by the states 
  constructed in \cite{Bonelli:2011aa,Gaiotto:2012sf} which are considered in section \ref{sec:Virasoro}, are different:
  the former state $\left| G_m \right>$ is specified by 
    \bea
    L_1 \left| G_m \right>
     =     \Lambda^{\frac{2}{m}} v_{1} \left| G_m \right>, 
           ~~~~
    L_m \left| G_m \right>
     =     \Lambda^2 \left| G_m \right>,
           \label{G}
    \eea
  and is not an eigenstate for $L_k$ with $1 < k < m$.
  The latter state $\left| I_n \right>$ is specified by
    \bea
    L_{n} \left| I_n \right>
     =     \ell_{n} \left| I_n \right>, 
           ~~ \ldots, ~~
    L_{2n} \left| I_n \right>
     =     \ell_{2n} \left| I_n \right>,
           \label{Gtilde}
    \eea
  where $\ell_{k}$ ($n \leq k \leq 2n$) are constants,
  and is not an eigenstate for $L_k$ with $k<n$.
  
  In \cite{Bonelli:2011aa}, an explicit solution to the conditions \eqref{G} has been given:
    \bea
    \left| G_m \right>
    &=&    \sum_{\ell=0}^\infty \sum_{\ell_p}
           \Lambda^{2 \ell/m} \prod_{i=1}^{[\frac{m}{2}]} c_i^{\ell_{m-i}} 
           \prod_{a=1}^{[\frac{m-1}{2}]} v_a^{\ell_a}
           Q^{-1}_\Delta (m^{\ell_m} (m-1)^{\ell_{m-1}} \cdots 2^{\ell_2} 1^{\ell_1}; Y)
           L_{-Y} \left| \Delta \right>,
           \nonumber \\
    & &    \label{gn}
    \eea
  with $\ell$ is a level $\ell = \sum_{s=1}^{m} s \ell_s$.
  Note that we assumed that the coefficient of the primary state of the expansion of the irregular state is 1, namely
  $\left| G_{m} \right> = \left| \Delta \right> + \CO(\Lambda)$, where
  $\CO(\Lambda)$ terms include various descendant states.
  (As discussed above, if we allow the other normalization of the primary state, 
  the expansion of the irregular state could be different from that of \eqref{gn}. 
  But, we do not see this possibility in this Appendix.)
  We will see below that this state satisfies the conditions \eqref{G} in the convention of descendant fields:
  $L_{-k_{1}} L_{-k_{2}} \cdots \left| \Delta \right>$ with 
    \bea
    k_{1} \leq k_{2} \leq \ldots.
    \label{I-k}
    \eea 
  Furthermore, we will see that in the different convention of descendant fields:
    \bea
    k_{1} \geq k_{2} \geq \ldots,
    \label{II-k}
    \eea
  the explicit state \eqref{gn} satisfies the conditions \eqref{Gtilde} for $2n = m$.

\subsection{Irregular states in the first convention}
  First of all, let us check that the state \eqref{gn} is indeed a solution of \eqref{G}
  in the first convention \eqref{I-k}.
  Note that the state \eqref{gn} satisfies
    \bea
    \left< \Delta \right| L_{Y} \left| G_m \right>
     =     \Lambda^{2\ell/m} \prod_{i=1}^{[\frac{m}{2}]} c_i^{\ell_{m-i}} 
           \prod_{a=1}^{[\frac{m-1}{2}]} v_a^{\ell_a}, ~~~~{\rm for}~
    Y
     =     m^{\ell_{m}} (m-1)^{\ell_{m-1}} \cdots 2^{\ell_{2}} 1^{\ell_{1}}.
           \label{inner}
    \eea
  In \cite{Bonelli:2011aa}, it was shown that this state satisfies (\ref{G}) and
    \bea
    L_k \left| G_m \right>
     =     0 ~~~{\rm for}~k > m.
    \eea
  For $L_k$ with $1 < k < m$, we obtain
    \bea
    L_{m-1} \left| G_m \right>
    &=&    \Lambda^{2(m-1)/m} \left( c_1 + (2 - m) \frac{\partial}{\partial v_{1}} \right)
           \left| G_m \right>,
           \label{Ln-1action} \\
    L_{m-2} \left| G_m \right>
    &=&    \Lambda^{2(m-2)/m} \left( c_2 + (3 - m) c_1 \frac{\partial}{\partial v_{1}}
         + \frac{(2-m)(3-m)}{2} \frac{\partial^2}{\partial v_{1}^2}
         + (4-m) \frac{\partial}{\partial v_{2}} \right)
           \left| G_m \right>,
           \nonumber 
    \eea
  and so on.
  A generic feature is that the action of $L_{m-k}$ starts with a term with $c_{k}$  
  and the remaining terms, although involved, can be written as differential operators in the parameters.
   
  For instance, the state $\left| G_4 \right>$ is given by
    \bea
    \left| G_4 \right>
     =     \sum_{\ell=0}^\infty \sum_{\ell_p}
           \Lambda^{\ell/2} c_1^{\ell_{3}} m^{\ell_{2}} v_1^{\ell_1}
           Q^{-1}_\Delta (4^{\ell_4} 3^{\ell_{3}} 2^{\ell_2} 1^{\ell_1}; Y)
           L_{-Y} \left| \Delta \right>,
    \eea
  where we have renamed $c_{2}$ as $m$ which corresponds to the dimension-one mass parameter of the gauge theory.
  This state satisfies
    \bea
    L_1 \left| G_4 \right>
    &=&    \Lambda^{\frac{1}{2}} v_1 \left| G_4 \right>, ~~~
    L_2 \left| G_4 \right>
     =     \Lambda \left( m - c_1 \frac{\partial}{\partial v_{1}}
         + \frac{\partial^2}{\partial v_{1}^2} \right) \left| G_4 \right>,
           \nonumber \\
    L_3 \left| G_4 \right>
    &=&    \Lambda^{\frac{3}{2}} \left( c_1 - 2 \frac{\partial}{\partial v_1} \right) \left| G_4 \right>,
           ~~~
    L_4 \left| G_4 \right>
     =     \Lambda^2 \left| G_4 \right>.
           \label{Gaiotto4}
    \eea

\subsection{Irregular states in the second convention}
  Let us next consider the same state \eqref{gn} in the convention (\ref{II-k}).
  Let us below see that when $m=2n$, this state satisfies the same conditions as those of $\left| I_n \right>$.
  When $m=2n$, we have a state
    \bea
    \vert \widetilde{G_{2n}} \rangle
    &=&    \sum_{\ell=0}^\infty \sum_{\ell_p}
           \Lambda^{\ell/n} \prod_{i=1}^{n-1} c_i^{\ell_{2n-i}} 
           v_i^{\ell_i} m^{\ell_{n}}
           Q^{-1}_\Delta (2n^{\ell_{2n}} (2n-1)^{\ell_{2n-1}} \cdots 2^{\ell_2} 1^{\ell_1}; Y)
           L_{-Y} \left| \Delta \right>,
           \nonumber \\
    & &
           \label{gntildeeven}
    \eea
  where we have renamed $c_{k}$ as $m$, as in previous section.
  This state satisfies 
    \bea
    \langle \Delta | L_{Y} |\widetilde{G_{2n}} \rangle
     =     \Lambda^{\ell/n} \prod_{i=1}^{n-1} c_i^{\ell_{2n-i}} 
           v_i^{\ell_i} m^{\ell_{n}}, ~~~~{\rm for}~
    Y
     =     1^{\ell_{1}} 2^{\ell_{2}} \cdots 2n^{\ell_{2n}}.
    \eea
  Note that $Y$ is different from (\ref{inner}).
  
  One can check that
    \bea
    \langle \Delta | L_{Y} L_{2n-s} | \widetilde{G_{2n}} \rangle
    =     \Lambda^{\frac{2n-s}{n}} c_{s} \langle \Delta | L_{Y} | \widetilde{G_{2n}}\rangle,
    \eea
  for $0 \leq s < n$, with $c_{0} = 1$, and
    \bea
    \langle \Delta | L_{Y} L_{n} | \widetilde{G_{2n}} \rangle
         =     \Lambda m \langle \Delta | L_{Y} | \widetilde{G_{2n}} \rangle.
    \eea
  This means 
    \bea
    L_{2n-s} \widetilde{\left| G_{2n} \right>}
    &=&    \Lambda^{\frac{2n-s}{n}} c_{s} | \widetilde{G_{2n}} \rangle
            ~~~~ {\rm for}~~ 0 \leq s < n,
           \nonumber \\
    L_{n} | \widetilde{G_{2n}} \rangle
    &=&    \Lambda m |\widetilde{G_{2n}}\rangle,
    \eea
  For $L_{s}$ ($s<n$), the state is not the eigenstate, 
  but acts as differential operators with respects to $c_{i}$ parameters,
  {\it e.g.}, one can check that 
    \bea
    \langle \Delta | L_{Y} L_{n-1} | \widetilde{G_{2n}} \rangle
     =     \Lambda^{\frac{n-1}{n}} \left( v_{n-1} + 2\frac{\partial}{\partial c_{n-1}}  
         + c_{1} \frac{\partial}{\partial m} \right) \langle \Delta | L_{Y} \widetilde{G_{2n}} \rangle,
    \eea
  which means that
    \bea
    L_{n-1} |\widetilde{G_{2n}} \rangle
     =     \Lambda^{\frac{n-1}{n}} \left( v_{n-1} + 2\frac{\partial}{\partial c_{n-1}}  
         + c_{1} \frac{\partial}{\partial m} \right) |\widetilde{G_{2n}} \rangle.
    \eea

  For instance, the state $|\widetilde{G_4} \rangle$ is given by
    \bea
    L_{4} |\widetilde{G_4} \rangle
    &=&    \Lambda^{2} |\widetilde{G_4} \rangle, ~~~
    L_{3} |\widetilde{G_4} \rangle
     =     \Lambda^{\frac{3}{2}} c_{1} \widetilde{\left| G_4 \right>}, ~~~
    L_{2} |\widetilde{G_4} \rangle
     =     \Lambda m |\widetilde{G_4} \rangle,
           \nonumber \\
    L_{1} |\widetilde{G_4} \rangle
    &=&    \Lambda^{\frac{1}{2}} \left( v_{1} + 2\frac{\partial}{\partial c_{1}}
         + c_{1} \frac{\partial}{\partial m} \right) |\widetilde{G_4} \rangle.
    \eea
  We can check that these are exactly the same as the condition satisfied by the state $\left| I_{2} \right>$.
  Indeed, the state $\left| I_{2} \right>$ is specified by
    \bea
    L_{4} \left| I_{2} \right>
    &=&    \hat{c}_{2}^{2} \left| I_{2} \right>, ~~~
    L_{3} \left| I_{2} \right>
     =     2\hat{c}_{1} \hat{c}_{2} \left| I_{2} \right>, ~~~
    L_{2} \left| I_{2} \right>
     =     (2 \hat{c}_{0} \hat{c}_{2} + \hat{c}_{1}^{2}) \left| I_{2} \right>,
           \nonumber \\
    L_{1} \left| I_{2} \right>
    &=&    \hat{v}_{1} + 2 \hat{c}_{0} \hat{c}_{1}
         + \hat{c}_{2} \frac{\partial}{\partial \hat{c}_{1}} \left| I_{2} \right>,
    \eea
  where we used hatted variables for the parameters in section \ref{sec:Virasoro}.
  (We ignored the terms including $Q$.)
  The first three equations implies the relations among the parameters
    \bea
    \hat{c}_{2} 
     =     \Lambda, ~~~~
    \hat{c}_{1}
     =     \frac{c_{1} \Lambda^{1/2}}{2}, ~~~~
    \hat{c}_{0}
     =     \frac{1}{2} (m - \frac{c_{1}^{2}}{4}).
    \eea
  Thus, the derivative can be written in terms of the parameters of our state as
    \bea
    \hat{c}_{2} \frac{\partial}{\partial \hat{c}_{1}}
     =     \Lambda^{\frac{1}{2}} \left( 2\frac{\partial}{\partial c_{1}}
         + c_{1} \frac{\partial}{\partial m} \right).
    \eea
  This shows the equivalence of our state and the one in \cite{Gaiotto:2012sf},
  with the identification $v_{1} = \hat{v}_{1} + 2 \hat{c}_{0} \hat{c}_{1}$.


\section{Irregular states of $U(1)$ current algebra}

\label{sec:U(1)}

In this section we want to show that in the case of $U(1)$ current algebra the irregular states
obtained by the confluence of the vertex operators are nothing but the standard coherent 
states. The crucial point here is the free field nature of the $U(1)$ current algebra.
Namely all the positive modes $a_n~(n >0)$ of the $U(1)$ current are mutually 
commuting and the Verma module is the (infinite) tensor product of the Fock space
of a single harmonic oscillator.

Let us introduce a free chiral boson;
\beq
\varphi(z) = q + a_0 \log z - \sum_{ n \neq 0} \frac{a_n}{n} z^{-n},
\eeq
with the commutation relations:
\beq
[ a_m, q ] = \delta_{m, 0}, \qquad [ a_m, a_n ] = m~\delta_{m+n, 0}. 
\eeq
Associated $U(1)$ current is
\beq
J(z) = \partial\varphi(z) = \sum_{n \in \mathbb{Z}} a_n~z^{-n-1}
\eeq
and the Fock vacuum $\vert \alpha \rangle = V_\alpha(z) \vert 0 \rangle$ is created by
the vertex operator
\beq
V_\alpha (z) =~: e^{~\alpha\varphi(z)} :
\eeq
By the OPE
\beq
J(z) V_\alpha(w) \sim \frac{\alpha}{z-w} V_\alpha(w),
\eeq
we have the following action on $\vert R_n \rangle 
= V_{\alpha_1} (z_1) \cdots V_{\alpha_n} (z_n) \vert \alpha_0 \rangle$:
\bea
J(y) \vert R_n \rangle &=& \left( \frac{\alpha_1}{y -z_1} + \cdots 
+ \frac{\alpha_n}{y - z_n} + \frac{\alpha_0}{y}\right) \vert R_n \rangle \\
&=& \frac{P_n(y)}{y \prod_{i=1}^n (y - z_i)} \vert R_n \rangle,
\eea
where
$P_n(y) = c_0 y^n + c_1 y^{n-1} + \cdots + c_{n-1} y + c_n$. 
Then we will take the limit $z_i \to 0$ and $\alpha_i \to
\infty$, while keeping $c_0, c_1, \ldots, c_n$ finite. 
The limit state
$\vert I_n \rangle_F = \lim_{z_i \to 0, \alpha_i \to \infty} \vert R_n \rangle$ satisfies;
\beq
J(y) \vert I_n \rangle_F = \sum_{i=0}^n \frac{c_i}{y^{1+i}} \vert I_n \rangle_F.
\eeq
This means that
\beq
a_i \vert I_n \rangle_F = c_i \vert I_n \rangle_F~~(1 \leq i \leq n), \qquad a_k \vert I_n \rangle_F =0,~~( k \geq n),
\eeq
and hence $\vert I_n \rangle_F$ is nothing but the standard coherent state
with eigenvalue $c_i$ for the $i$-th oscillator mode. 
The most simple example is the collision of two vertex operators, where 
\beq
P_1(y) = c_0 y + c_1, \qquad c_0 = \alpha_0 + \alpha_1, \quad c_1 = - \alpha_0 z_1
\eeq
We take the limit $z_1 \to 0$ and $\alpha_0 \to +\infty$, 
keeping $\alpha_0 + \alpha_1$ and $c_1$ finite (or $\alpha_1 \to -\infty$). 
This is a point like limit of the dipole with an infinite charge. 
The coherent state produced by the confluence of $(n+1)$ punctures 
can be generated by the generalized vertex operator on the primary state;
as
\beq
\vert I_n ; c_i, \alpha \rangle_F
= \lim_{z \to 0} \exp \left(\sum_{k=1}^n \frac{1}{k} (c_k~\partial^k \varphi(z)  \right)  \vert \alpha \rangle.
\eeq


\bibliographystyle{ytphys}
\small\baselineskip=.97\baselineskip
\bibliography{ref}

\end{document}